\documentclass{pasj00}
\usepackage{natbib,aas_macros}
\newcommand{\bm}[1]{\mbox{\boldmath$#1$}}
\newcommand{\cd}[2]{{\rm CD}#1\_#2}
\newcommand{\A}[2]{{\rm A}\_{#1}\_{#2}~}
\newcommand{\B}[2]{{\rm B}\_{#1}\_{#2}~}
\newcommand{\pv}[2]{{\rm PV}{#1}\_{#2}~}

\begin{document}
\SetRunningHead{T. Hamana et al.}{Cosmological constraints from
  Subaru weak lensing cluster counts}
\Received{}
\Accepted{}
\Published{\today}

\title{Cosmological constraints from Subaru weak
  lensing cluster counts}

\author{Takashi \textsc{Hamana}\altaffilmark{1}, 
Junya \textsc{Sakurai}\altaffilmark{1},
Michitaro \textsc{Koike}\altaffilmark{1},
Lance \textsc{Miller}\altaffilmark{2}}
\altaffiltext{1}{National Astronomical Observatory of Japan, Mitaka, 
Tokyo 181-8588, Japan}
\altaffiltext{2}{Department of Physics, Oxford University, Keble Road,
  Oxford OX1 3RH, UK} 

\KeyWords{cosmology: observations --- dark matter --- cosmological
  parameters --- large-scale structure of universe }

\maketitle

\begin{abstract}
We present results of weak lensing cluster counts obtained from 11
degree$^2$ Subaru/SuprimeCam data.
Although the area is much smaller than previous work dealing with weak
lensing peak statistics, the number density of galaxies usable for weak
lensing analysis is about twice as large as those.
The higher galaxy number density reduces the noise in the weak lensing
mass maps, and thus increases the signal-to-noise ratio ($S/N$) of peaks
of the lensing signal due to massive clusters. This enables us to construct
a weak lensing selected cluster sample by adopting
a high threshold $S/N$, 
such that the contamination rate due to false signals is small.
We find 6 peaks with $S/N>5$.
For all the peaks, previously identified clusters of galaxies are
matched within a separation of 1 arcmin, demonstrating good correspondence
between the peaks and clusters of galaxies. 
We evaluate the statistical error in the weak lensing cluster counts
using mock weak lensing data generated from full-sky ray-tracing simulations, 
and find $N_{\rm peak}= 6 \pm 3.1$ in an effective area of 9.0 degree$^2$.
We compare the measured weak lensing cluster counts with the
theoretical model prediction based on halo models and place
the constraint on $\Omega_m-\sigma_8$ plane which is found to be consistent
with currently standard $\Lambda$CDM models.
It is demonstrated that the weak lensing cluster counts can place a
unique constraint on $\sigma_8-c_0$ plane, where $c_0$ is the
normalization of the dark matter halo mass--concentration relationship.
Finally we discuss prospects for ongoing/future wide field optical
galaxy surveys.
\end{abstract}

%
%
\section{Introduction}
The $\Lambda$-flat universe with cold dark matter ($\Lambda$CDM) is now
considered as the standard theoretical framework for cosmic structure formation.  
In the $\Lambda$CDM paradigm, dark matter haloes of a galaxy and
a cluster of galaxies form though the hierarchical bottom-up assembly of
smaller structures. 
The averaged density profile of the dark matter haloes found in $N$-body
simulations has a universal broken power-law form
\citep{1997ApJ...490..493N}. 
The universal profile is characterized by a single parameter (called the
concentration parameter) describing the ratio of the scale radius (where
the power-law slope is approximately $-2$) to the virial radius which
defines a halo's mass. 
The concentration parameter is known to be related to the halo mass,
a correlation which most likely arises from the mass assembly history of
haloes \citep{2001MNRAS.321..559B,2009ApJ...707..354Z,2014MNRAS.441..378L,2015ApJ...799..108D}.
Since the mass assembly history depends on the cosmological model, the
mass--concentration relation does as well
\citep{2008MNRAS.391.1940M,2013ApJ...768..123K}. 
Therefore placing constraints on both the cosmological model and the
mass--concentration relation simultaneously allows us to test the 
$\Lambda$CDM structure formation scenario.

Clusters of galaxies serve as one of the most important tests of the
cosmic structure formation model. 
Due to their high virial temperatures, the dissipative cooling of
baryons is less effective, and thus the density profile of clusters may
be well approximated by the dark matter distribution, allowing us to
test the mass--concentration relation
\citep[see for a recent review][and references therein]{2013SSRv..177..119E}.
Also, cluster number counts have placed useful constraint on the
cosmological parameters
\citep{2009ApJ...692.1060V,2010MNRAS.406.1759M,2010ApJ...708..645R}.

It is argued in recent studies that number counts of clusters of galaxies
identified in weak lensing mass maps (which we call weak lensing cluster
counts) can be a powerful probe of both the cosmological model and the
mass--concentration relation \citep{2014JCAP...08..063M,2014arXiv1409.5601C}.
This can be simply understood as follows:
First, the number density of massive clusters of galaxies, which
appear as high peaks on the weak lensing mass maps, depends of the
cosmological model \citep{2001ApJ...561...13B,2004ApJ...609..603H}.
Second, peak heights depend on the density profile of
clusters and thus on the mass--concentration relation \citep{2011MNRAS.416.2539K,2012MNRAS.425.2287H}. 

Both observational techniques and theoretical models of weak lensing
cluster counts have been greatly developed in the last decade.
On the observational side, systematic weak lensing cluster searches have
become practicable
\citep{2002ApJ...580L..97M,2007ApJ...669..714M,2006ApJ...643..128W,2007A&A...462..459G,2007A&A...462..875S,2009ApJ...702..980K,2011MNRAS.413.1145B,2012ApJ...748...56S}.
\citet{2007ApJ...669..714M} presented the first large
sample of weak lensing peaks (which are plausible candidates of
clusters/groups of galaxies) in $16.7$ degree$^2$ Subaru weak lensing
survey data, with 100 peaks with signal-to-noise
ratio exceeding $S/N>3.7$. 
\citet{2012ApJ...748...56S} reported 301 weak lensing high peaks ($S/N>3.5$) located from
64 degree$^2$ of CFHTLS-wide data, among those peaks, they confirmed
85 groups/clusters. 
On the theoretical side, models of the weak lensing cluster counts
have been developed: Based on the pioneering work by
\citet{2001A&A...378..361B}, some improvements have been made, including
the  effect of the galaxy intrinsic ellipticities
\citep{2004MNRAS.350..893H,2010ApJ...719.1408F}, the effect of the
large-scale structures \citep{2010ApJ...709..286M,2010A&A...519A..23M}, the
effect of the diversity of the dark matter distribution within clusters
\citep{2012MNRAS.425.2287H},  
and an optimal choice of the filter function \citep{2005ApJ...624...59H,2005A&A...442..851M}.
Now, the theoretical models were tested  (or calibrated) against $N$-body
simulations and good agreements were found (see the above
references).

In fact, weak lensing peak counts have now become a practical tool for
investigating cosmological models.
\citet{2014arXiv1412.0757L} used weak lensing peak counts combined with
weak lensing power spectrum measured from the CFHTLenS data to constrain
cosmological parameters. They used not only high peaks but also low
peaks ($-0.04 <\kappa < 0.12$) in order to efficiently utilize cosmological
information contained in the weak lensing mass map.
They found that the constraints from peak counts are comparable to
those from the cosmic shear power spectrum.
\citet{2014arXiv1412.3683L} used weak lensing peak counts within $3<S/N<6$ from
$\sim 130$ degree$^2$ of the Canada-France-Hawaii Telescope Stripe82
Survey data. They demonstrated the capability to
put a constraint on the mass--concentration relation together with the
cosmological parameters from weak lensing peak counts.

In this paper, we examine weak lensing cluster counts obtained from 
11 degree$^2$ of weak lensing mass maps generated from deep
Subaru/SuprimeCam data.
In contrast to previous work \citep{2014arXiv1412.0757L,2014arXiv1412.3683L}, we
deal only with high peaks, with $S/N>5$.
Such high peaks mostly associate with a real massive cluster of
galaxies, and thus we shall use the term ``weak lensing cluster counts''
instead of ``weak lensing peak counts''.
Our higher threshold enables us to make a reliable theoretical prediction
of the weak lensing cluster counts utilizing models of dark matter haloes
which are well calibrated with $N$-body simulations.
For the first time, we use the weak lensing cluster counts to place
cosmological parameters and the mass--concentration relationship.
In doing this, we explore, in an experimental manner, the capability
and current issues of weak lensing cluster counts for cosmological studies.

The structure of this paper is as follows. 
In Section~\ref{sec:dataanalysis}, we summarize the Subaru/SuprimeCam data,
data processing, and weak lensing shape measurement.
We also examine the additive and multiplicative bias in the measured
galaxy ellipticities.
In section~\ref{sec:kapmap-peak}, we first describe 
construction of weak lensing mass maps.
Then we examine statistical properties of peaks paying special
attention to the residual systematics associated with data reduction
and/or weak lensing shear measurement. 
In that section, we also summarize the high $S/N$
peaks and their possible counterpart clusters of galaxies searched in
known cluster databases.
In Section~\ref{sec:constraints}, the measured weak lensing cluster
counts are compared with the theoretical prediction to place constraints on
the cosmological parameters and the dark matter halo mass--concentration
relationship.
Finally, summary and discussion are given in Section~\ref{sec:summary}.   
In Appendix \ref{appendix:sip2tpv}, the method for
conversion of WCSs from the astrometric SIP to TPV convention, which is
used in the present work, is described.

\section{Data analysis}
\label{sec:dataanalysis}

\subsection{Data set and data processing}
\label{sec:dataset}

%
%
\begin{table}
\caption{Summary of data sets. See Fig.~\ref{fig:snmaps} for
the coordinate and shape of each field}
\label{table:fields}
\begin{tabular}{lccc}
\hline
Field name & Area$^{a}$ & $n_g$$^{b}$ & Seeing$^{c}$ \\ 
\hline
XMM-LSS      & 3.6 (2.8) & 23.7 (20.9) & 0.51--0.71 \\
COSMOS       & 2.1 (1.6) & 28.8 (26.3) & 0.50--0.53 \\
Lockman-hole & 2.1 (1.6) & 25.5 (23.8) & 0.42--0.55 \\
ELAIS N1     & 3.6 (2.9) & 24.6 (22.0) & 0.45--0.69 \\
\hline
\end{tabular}\\
$^{a}$ Area after masking regions affected by bright stars in unit of
degree$^2$. The numbers in parentheses are the effective area ($A^{eff}$) after
cutting the edge regions within 1.5~arcmin from the boundary.\\
$^{b}$ The mean number density of galaxies used for the weak lensing analysis 
in the unit of arcmin$^{-2}$. The numbers in parentheses are the
effective number density defined by $n_g^{eff}=(\sum_i
w_i)^2/\sum_i w_i^2/A^{eff}$ \citep{2012MNRAS.427..146H}.\\
$^{c}$ Seeing full width at half-maximum in the unit of arcsec.
The smallest and largest values among the pointings are given.
\end{table}

We collected $i'$-band data taken with the Subaru/SuprimeCam
\citep{2002PASJ...54..833M} from the data archive {\it SMOKA}\footnote{\tt
  http://smoka.nao.ac.jp/} \citep{2002ASPC..281..298B}, 
under the following three conditions: data,
which are composed of a number of pointings, are contiguous at least 2
degree$^2$, the exposure time for each pointing is longer than 1800~sec,
and the seeing full width at half-maximum (FWHM) of each exposure is
better than 0.7 arcsec.
Four data sets meet these requirements and are summarized
in Table~\ref{table:fields}. 
Data of each pointing are composed of several dithered
exposures. Dithering scales are a few arcmin except for
the COSMOS field for which a specific dithering pattern was adopted
\citep[a combination of dithered pointings of the normal ``north is up''
  exposures and 90 degree rotated exposures, see ][]{2007ApJS..172....9T}.

\begin{figure}
\begin{center}
 \includegraphics[width=78mm]{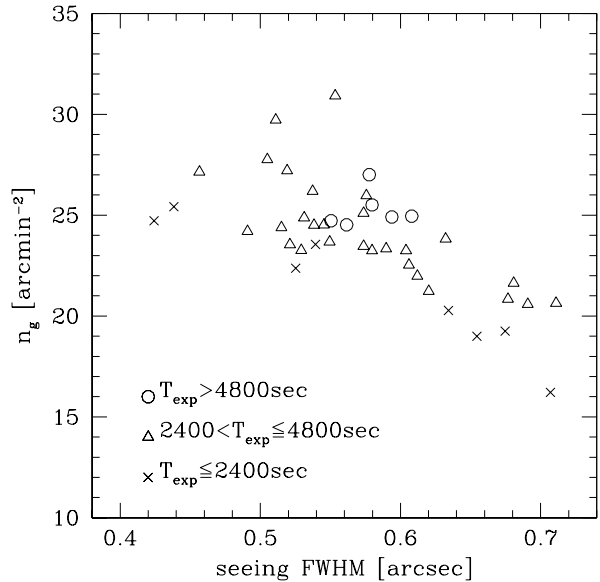}
\end{center}
\caption{The number densities of galaxies used for weak lensing analysis
  are plotted against the seeing FWHM for each pointing.
The different symbols represent the different exposure time. 
\label{fig:fig1.eps}}
\end{figure}

Each CCD dataset was reduced using an image analysis pipeline, {\it
  hscPipe}, developed for Hyper SuprimeCam \citep{2012SPIE.8446E..0ZM}
data (Furusawa et al. in prep.), which is based on {\it LSST-stack}
\citep{2008arXiv0805.2366I,2010SPIE.7740E..15A}, with being tuned for SuprimeCam data.
The mosaic stacking was done with {\it hscPipe} using the standard
correlation technique between star positions on CCD coordinates and
an external star catalog (we used the SDSS DR8),
and also between the star positions among different exposures.
We made stacked data on a pointing-to-pointing basis except for COSMOS
data for which a single stacked data was made by using {\it SCAMP}
\citep{2006ASPC..351..112B} and {\it SWarp} \citep{2002ASPC..281..228B}
from all the CCD data. 
Object detections were performed with {\it SExtractor}
\citep{1996A&AS..117..393B}.
Stars are selected in a standard way by
identifying the appropriate branch in the magnitude-object size (FWHM)
plane, along with the detection significance cut $S/N>10$. 
The number density of stars is found to be $\sim 1$~arcmin$^{-2}$ for
the four fields. We only use galaxies met the following three
conditions, (i) the detection significance of $S/N>8$, (ii) FWHM is
larger than the stellar branch, and (iii) the AB magnitude (for which we
adopted {\it SExtractor}'s {\tt MAG\_AUTO}) is in the
range of $22 <i'<25$.
The number density of resulting galaxy catalogue varies among pointings
due mainly to the variation in the seeing and weakly to
the exposure time (see Fig.~\ref{fig:fig1.eps}).

%
%

\subsection{Weak lensing shear estimate}
\label{sec:shear}

For the weak lensing shear estimates, we employ {\it lensfit}
\citep{2007MNRAS.382..315M,2008MNRAS.390..149K,2013MNRAS.429.2858M}, which is a
model-fitting method. We refer the reader to the above 
references for details of its method, implementation, and results from
tests on image simulations.
One important function of {\it lensfit} which should be noticed is that
galaxies on each individual exposure image are fitted by models that
have been convolved with the PSF for the image, and the resulting
likelihoods multiply to 
obtain the final combined likelihood.
This can avoid problematic issues in measuring weak lensing
shears from a stacked image; including deformation of object shapes by
interpolation of pixel data needed for resampling onto a co-added pixel
grid \citep{2007ApJS..172..203R,2008PASJ...60.1363H,2013MNRAS.429.2858M}, 
and a complex PSF
variation over a stacked image caused by
co-addition of different individual PSFs \citep{2013PASJ...65..104H}.  

Application of {\it lensfit} to SuprimeCam data was first done by
\citet{2012Natur.487..202D}.
We used the {\it lensfit} software suite which is designed for dealing
with SuprimeCam data.
For the actual implementation of {\it lensfit}, we basically follow one
developed for CFHTLenS \citep{2013MNRAS.429.2858M} and thus the
method is likelihood-based, and we adopt the same priors for the galaxy
model. 
In order to adjust {\it lensfit} to SuprimeCam data, we have made
following two modifications.  
(1) The size of postage-stamp image for galaxies was set to be
$40 \times 40$ ($48 \times 48$ for CFHTLenS) because of the
following two reasons: The first is the difference in the CCD pixel
sizes between SuprimeCam (0.202 arcsec) and MegaCam on CFHT (0.185
arcsec), and second is that more smaller galaxies are
detected in SuprimeCam data owing to their deeper image depth.
(2) The definition of a boundary of objects was modified in the
following way. In the analysis of CFHTLenS
\citep{2013MNRAS.429.2858M}, $2\sigma$ isophote was adopted
for a boundary of objects in the following two processes;
(I) masking neighbour objects in the same postage-stamp image, 
and (II) clipping of possible blending objects identified by the
overlapping of isophotes of close objects and by a mismatch between the
peak and centroid positions. 
Since the noise level of an image depends on the depth of data and our
data are much deeper than CFHTLenS data (thus smaller $\sigma$),
adopting the same $2\sigma$ isophote extends the object
boundary to outer skirt and significantly increases a chance a object 
to be classified as a blending object and to be rejected.
We thus decided to set the threshold isophote ($f_{\rm th}\sigma$) being the 
CFHTLenS-2$\sigma$ corresponding value; 
$f_{\rm th} = 2 \sqrt{T_{exp}/900{\rm sec}}$, where $T_{exp}$ is
the total exposure time for each pointing and $900{\rm sec}$ is the
SuprimeCam corresponding exposure time of the CFHTLenS ($3600{\rm sec}$).
Galaxies that are excluded in the process of {\it lensfit} shape
measurement is mostly due to the above criteria, plus a small fraction of
failures in the model-fitting step.
The fractions of excluded galaxies vary among pointings, and the average
value is 0.22.
This is comparable to that for the CFHTLenS ($\sim 0.2$),
supporting that our exposure-time dependent threshold results
in a similar level of screening of large, blended or complex galaxies.

In {\it lensfit}, the point spread functions (PSFs) are represented as 
postage-stamp data of pixel values measured from images of stars.
The PSFs are measured for each exposure.
The spatial variation of PSFs (pixel values of PSF postage-stamps) are
modeled by a two-dimensional polynomial function of position in the
SuprimeCam's field-of-view (FoV) that consists of a mosaic of $2\times
5$ CCD chips with low order coefficients being allowed to vary between
CCD chips. 
We employ a third-order polynomial for the FoV-based
model and a second-order for the chip-based model.
We did not find meaningful improvement using fourth-order for the FoV
model. 

%
%
\begin{figure}
\begin{center}
\includegraphics[width=84mm]{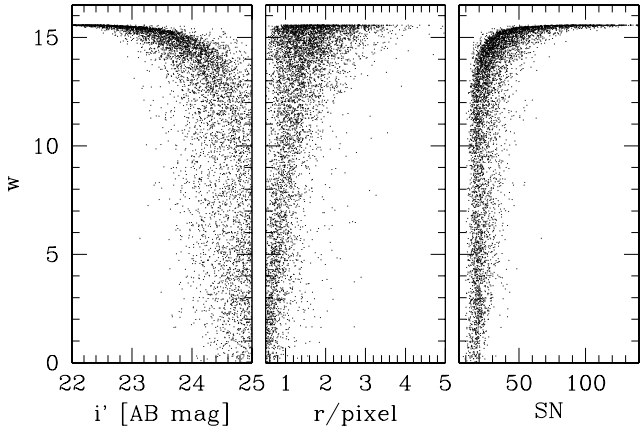}
\end{center}
\caption{The distribution of lensing weight for 8549 measured galaxies
  in one typical pointing, ELAIS\_02 (seeing FWHM $\sim 0.57$ arcsec and
  $T_{exp}=3600$ sec). The weights are shown as a function of ({\it
    left}) $i'$-band magnitude, ({\it center}) fitted semi-major axis
  disk scalelength, and ({\it right}) signal-to-noise ratio.
\label{fig:fig2.eps}}
\end{figure}

We adopted the same galaxy weighting scheme as one defined in
\citet{2013MNRAS.429.2858M}, which is according to the width of the
likelihood surface in model-fitting parameter space.
Fig.~\ref{fig:fig2.eps} shows an example of distribution of lensing
weights in one typical pointing, as a function of $i'$-band magnitude,
fitted semi-major axis disk scalelength, and signal-to-noise ratio.
Comparing with the corresponding plots for CFHTLenS (Fig.~2 of
\citet{2013MNRAS.429.2858M}), one may find very similar trends between
two analyses, confirming that the  weighting scheme properly works on
SuprimeCam data.

%
%
\begin{figure*}
\begin{center}
\includegraphics[height=74mm]{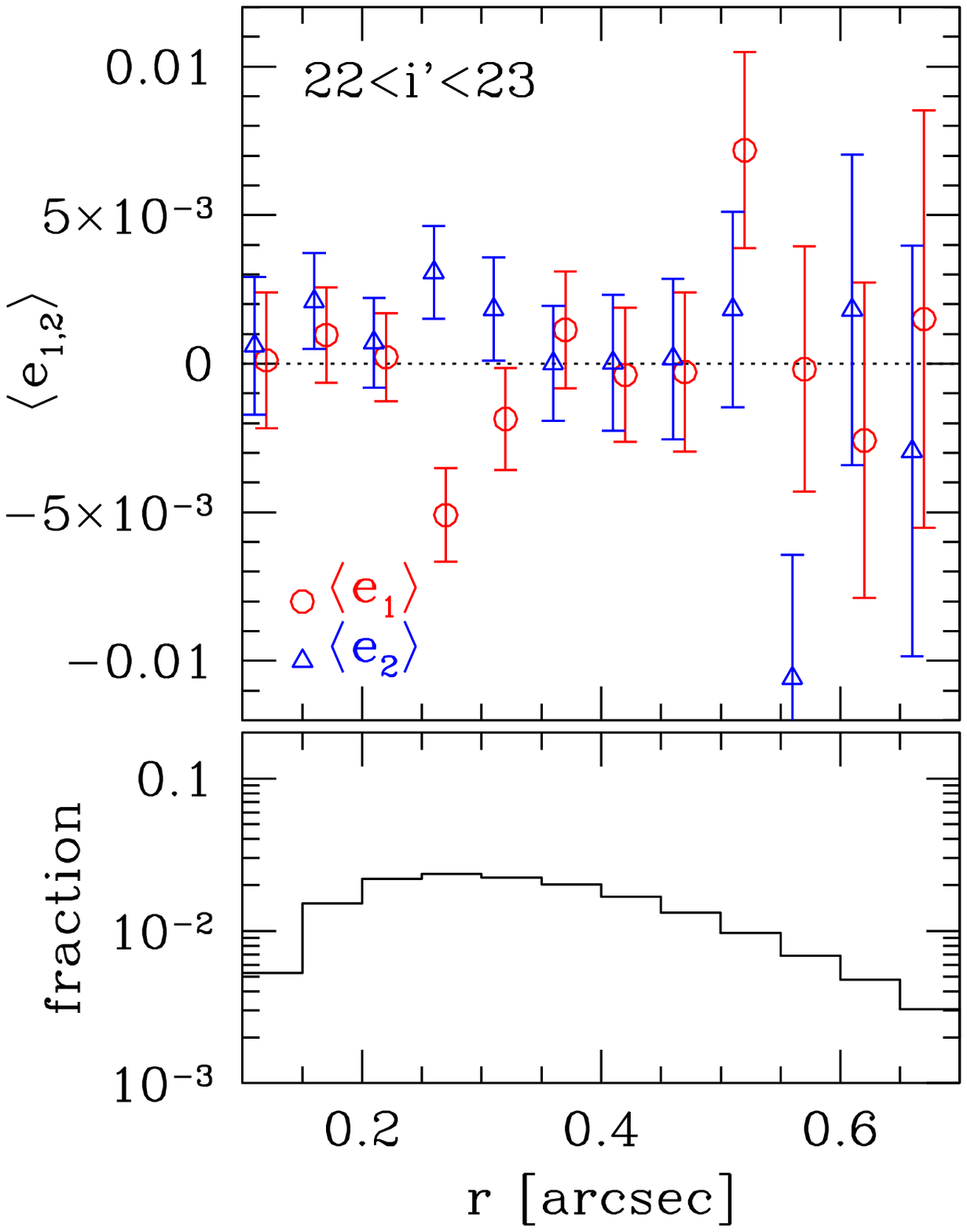}
\includegraphics[height=74mm]{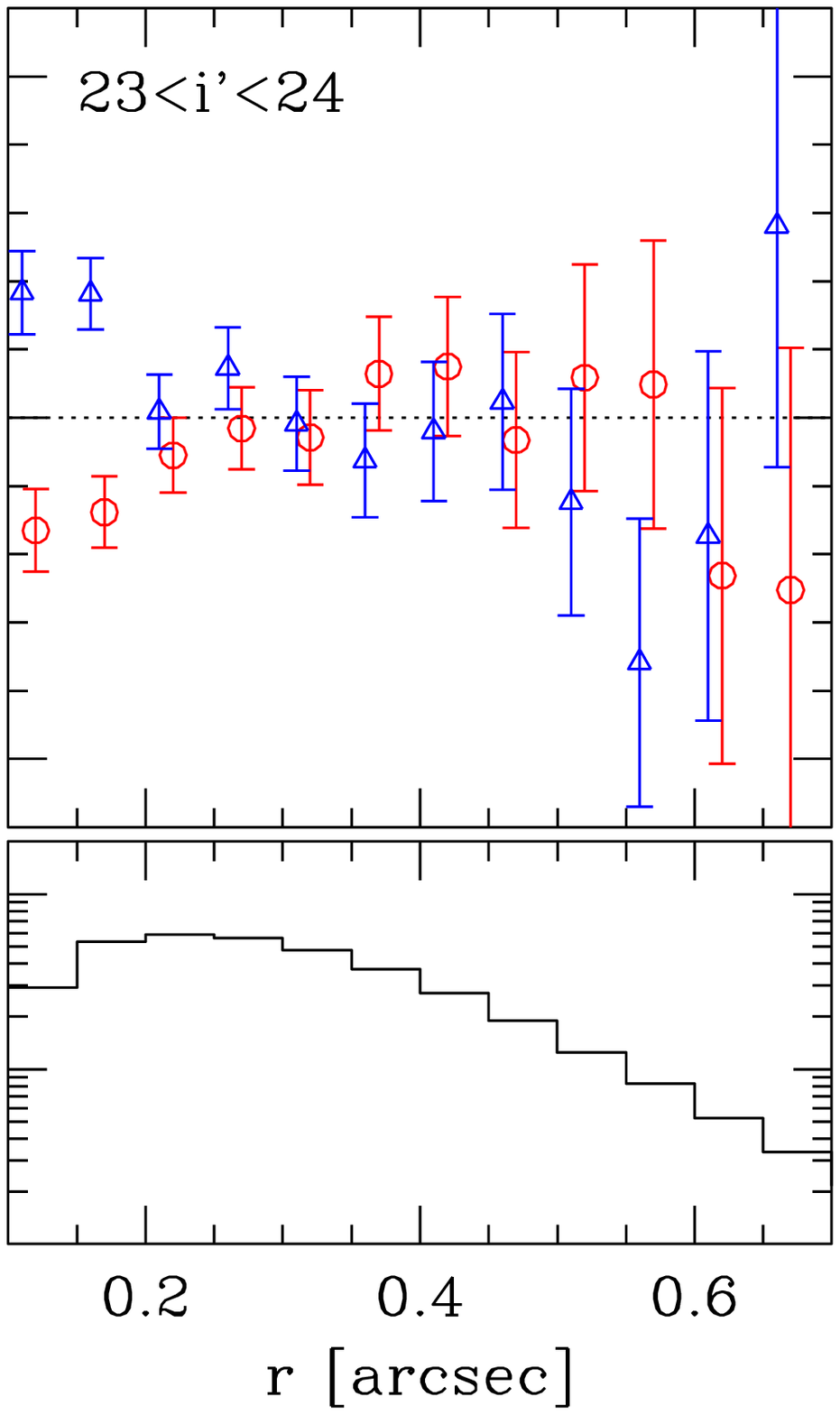}
\includegraphics[height=74mm]{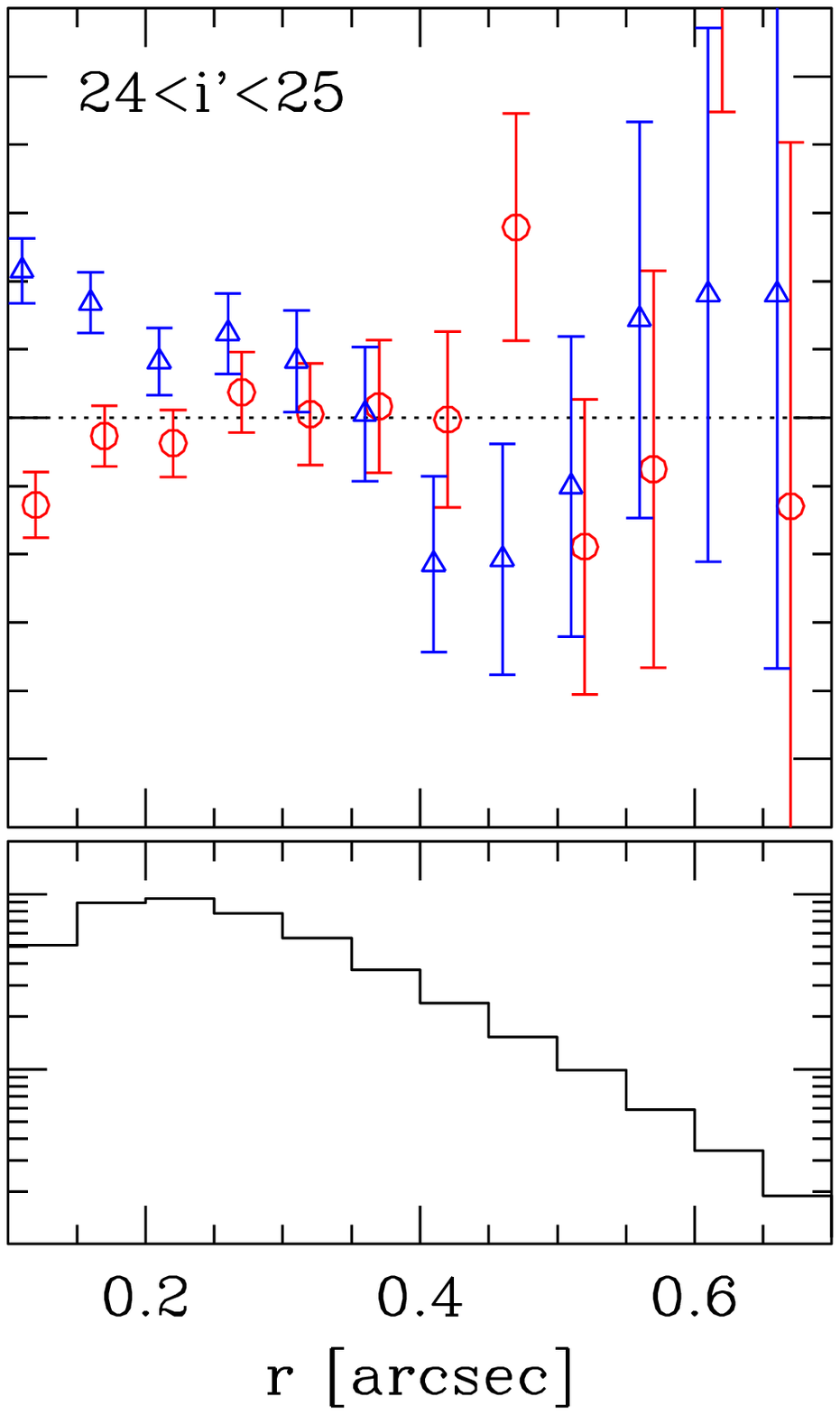}
\end{center}
\caption{{\it Top-panels}: Mean ellipticities of galaxies as a function
  of semi-major axis desk scalelength in three $i'$-band magnitude
  ranges. Open circles and triangles are for the $e_1$ and $e_2$
  component of the galaxy ellipticities, respectively,  obtained from
  {\it lensfit}  model-fitting. 
{\it Bottom-panels}: The weighted galaxy fraction for each
magnitude-scalelength grid.
\label{fig:fig3}}
\end{figure*}

It is known that galaxy shears estimated by {\it lensfit} are affected
by multiplicative and additive bias, which are usually represented by
\citep{2012MNRAS.427..146H}, 
$e^{\rm obs} = (1+m)[\gamma + e^{\rm int}]+c$.
\citet{2013MNRAS.429.2858M} used realistic image simulations to estimate
the multiplicative bias in CFHTLenS data, and found a size- and
$S/N$-dependent bias with an average value of 6 percent.
They derived the fitting function of the multiplicative bias [equation
(14) of \citet{2013MNRAS.429.2858M}] that we employ for the current
study.  Using the fitting function, we evaluated an weighted average
bias over source galaxies and found $\langle m \rangle =-0.059$.
It is not clear whether the fitting function that was derived using mock
CFHTLenS simulation data is applicable to the SuprimeCam data.
We may, however, say that this does not seriously affect our results for
the following two reasons:
First, we don't use a lensing induced quantity
(shear or convergence) directly, but use the signal-to-noise ratio map of the
lensing convergence (the convergence map normalized by its noise map).
Therefore, our primary observational quantity, the peak counts, is not
affected by the mutiplicative bias (see the next section). 
We, however, need to take the multiplicative bias into account
in making the theoretical prediction of the peak counts.
In order to do this, we will make a crude bias correction utilizing
the average value of $\langle m \rangle =-0.059$ (see \S \ref{sec:model}
for details). Since the statistical error in the peak counts is estimated as about
50 percent (see \S \ref{sec:error}), the uncertainty associated with the
multiplicative bias should be minor, that is the second reason.

Concerning the additive bias, we evaluated it in an empirical manner
using weighted mean value of ellipticities over galaxies and found
non-zero values with; 
$\langle c_1\rangle = \langle w e_1^{\rm obs}\rangle = -0.00046\pm
0.00030$, and
$\langle c_2\rangle = \langle w e_2^{\rm obs}\rangle = 0.0012\pm
0.00030$.
In order to look into details of this, we computed the additive bias as
a function of $i'$-band magnitude and
scalelength and found that the non-zero biases are mainly caused by
faint-small galaxies as is shown in Fig.~\ref{fig:fig3}.
This is in marked contrast to the result of CFHTLenS that the
small-bright galaxies were found to be a cause of the significant
additive bias in $\langle c_2\rangle$ \citep[][note that their $\langle
  c_1\rangle = 0.0001\pm 0.0001$ was found to be consistent with
  zero]{2012MNRAS.427..146H}. The cause of these non-zero bias is
unknown, but those different trends for different data sets may suggest
that it may be caused by any instrumental or data-dependent effect. 
\citet{2012MNRAS.427..146H} calibrated the bias in an empirical manner,
but we don't do it because of the following two reasons:
The first is that we search for high peaks in weak lensing mass
maps which are obtained from azimuthally averaged tangential galaxy
shears with an axisymmetric kernel. In that operation, the
additive bias is partly canceled out.
The second is that we deal with very high peaks corresponding to $\gamma
\gtrsim 0.08$, thus the additive bias with the above small amplitude does
not have a significant impact on our results.

\section{Weak lensing mass maps and high $S/N$ peaks}
\label{sec:kapmap-peak}

\subsection{Weak lensing mass maps}
\label{sec:kapmap}

%
%
\begin{figure*}
\begin{center}
\includegraphics[width=64mm,angle=-90]{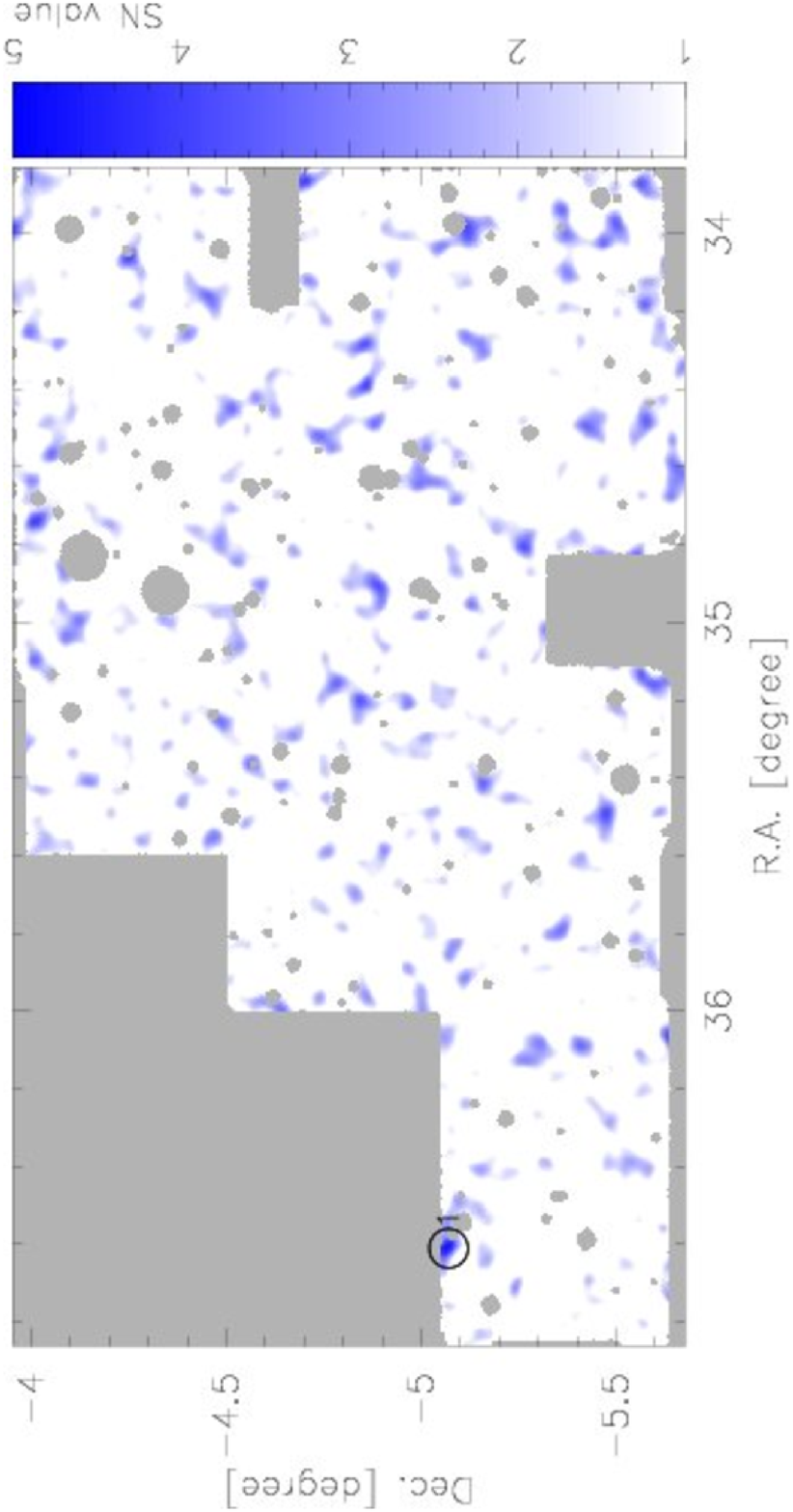}
\vspace{5mm}\\
\includegraphics[width=56mm,angle=-90]{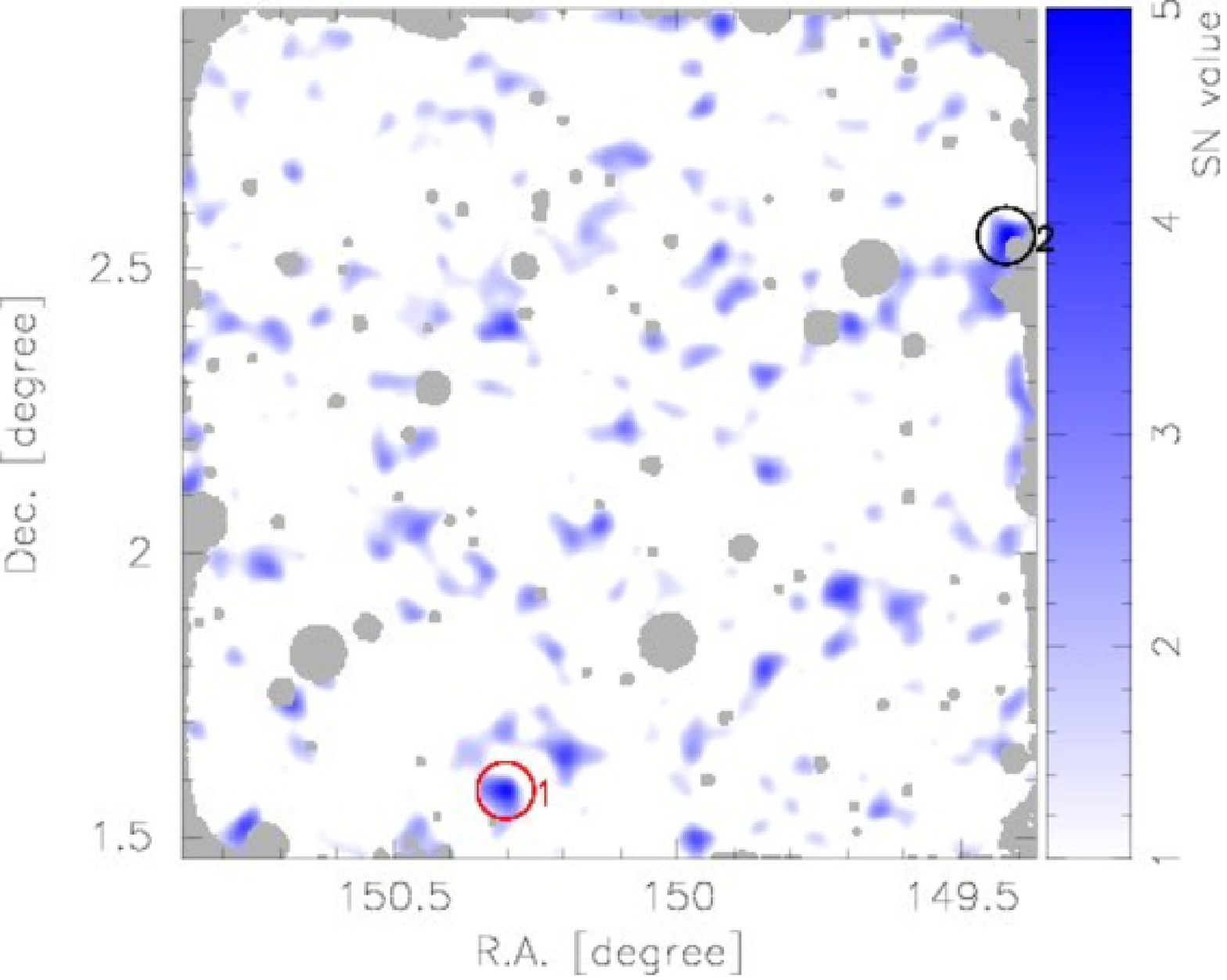}
\hspace{5mm}
\includegraphics[width=60mm,angle=-90]{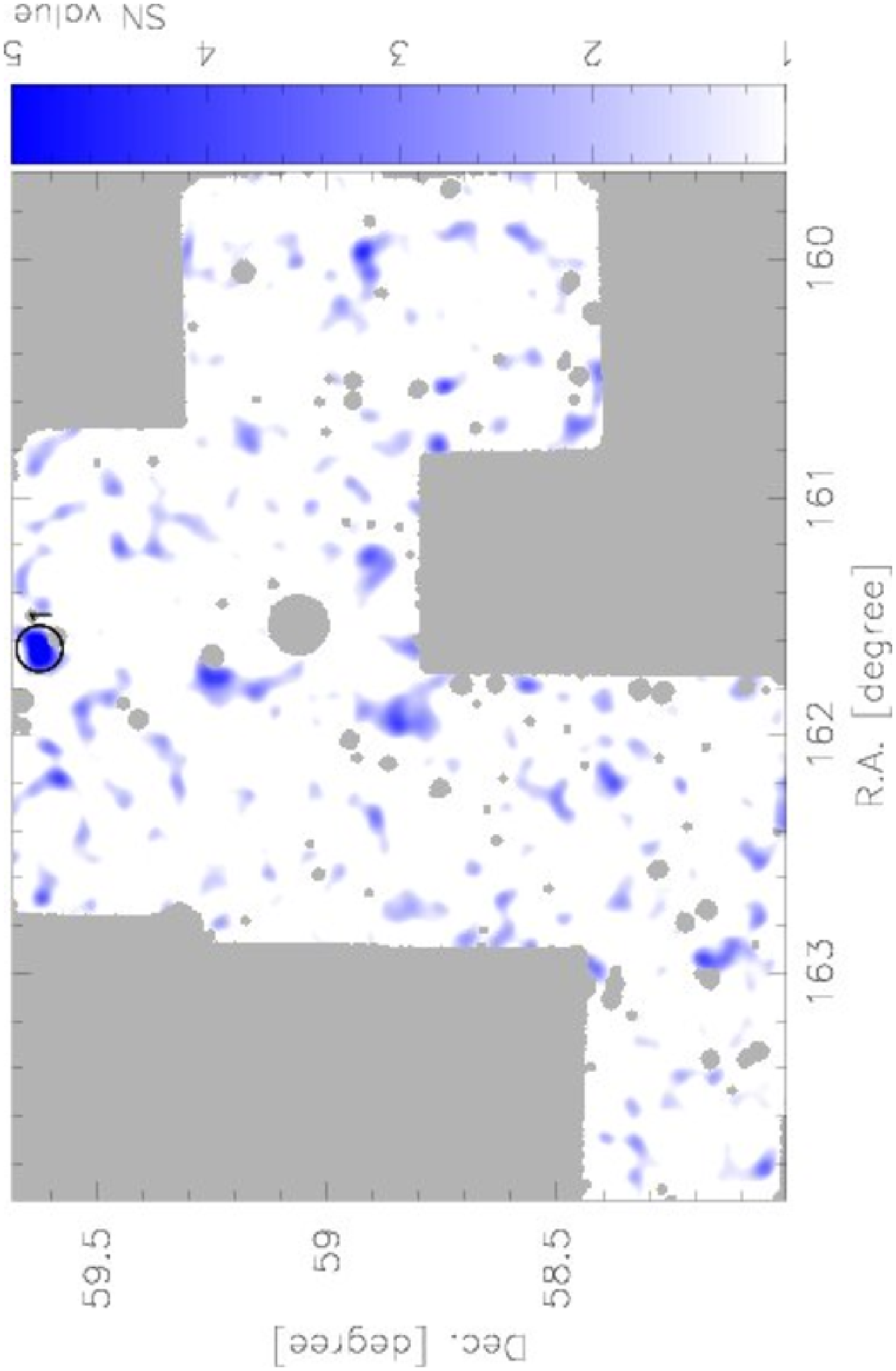}
\vspace{5mm}\\
\includegraphics[width=80mm,angle=-90]{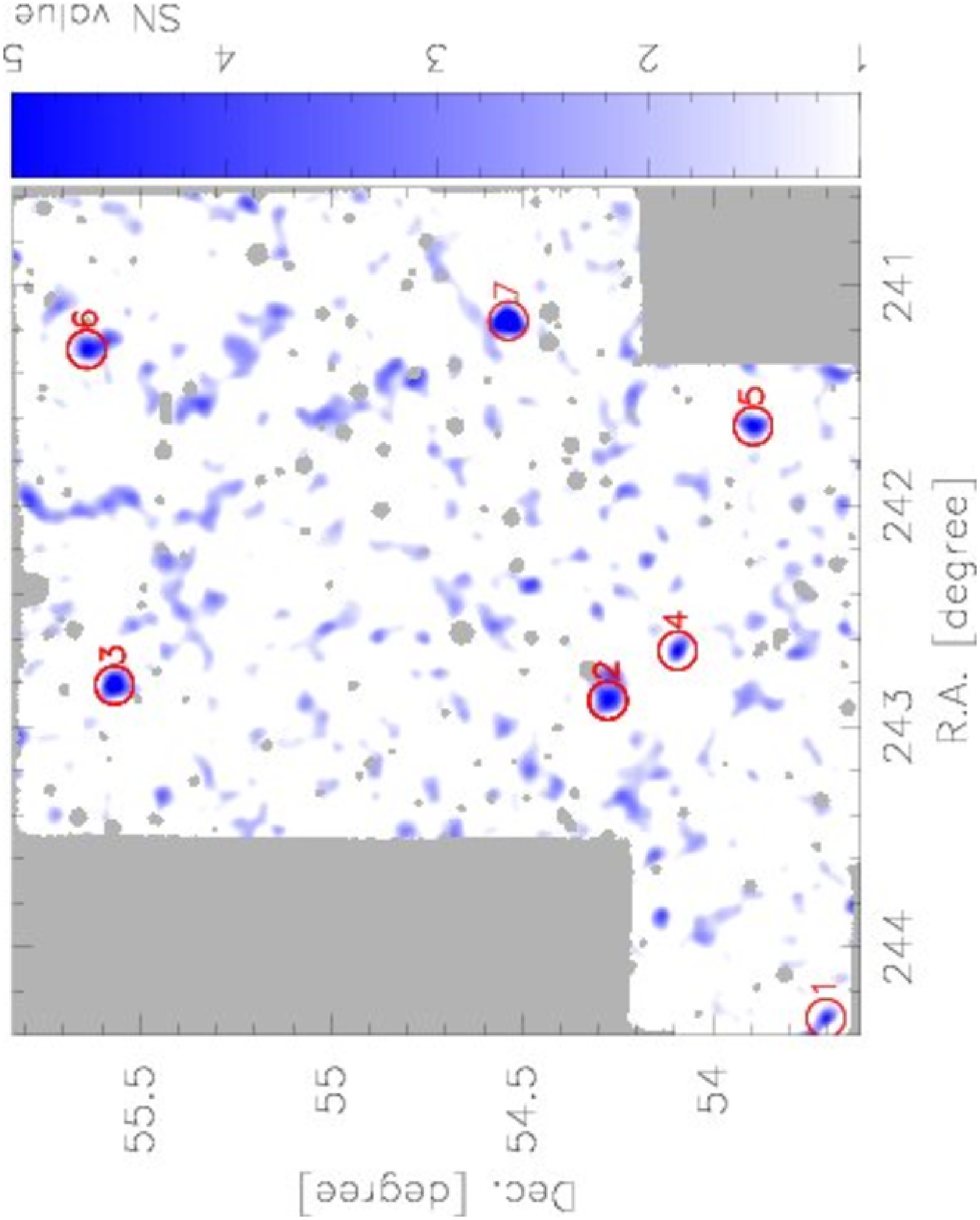}
\end{center}
\caption{$\cal K$ maps for the XMM-LSS ({\it top}), COSMOS ({\it middle-left}),
  Lockman-hole ({\it middle-right}), and ELAIS-N1 ({\it bottom}) are
  shown by blue scale. 
Gray regions are either outside of data regions or
masked regions where the data are affected by bright stars. 
High peaks with $S/N>4.5$ are marked with red circle along with ID
given in Table~ \ref{table:peaks}. Black circles are for high peaks
within the edge region.
\label{fig:snmaps}}
\end{figure*}

The weak lensing mass map which is the smoothed
lensing convergence field ($\kappa$) is evaluated from the tangential
shear data by \citep{1996MNRAS.283..837S}
\begin{equation}
\label{eq:shear2kap}
{\cal K }(\bm{\theta}) 
=\int d^2\bm{\phi}~ \gamma_t(\bm{\phi}:\bm{\theta}) Q(|\bm{\phi}|),
\end{equation}
where $\gamma_t(\bm{\phi}:\bm{\theta})$ is the tangential component of
the shear at position $\bm{\phi}$ relative to the point $\bm{\theta}$
and $Q$ is the filter function for which we adopt the truncated Gaussian
function (for $\kappa$ field) \citep{2012MNRAS.425.2287H},
\begin{equation}
\label{eq:Q}
Q(\theta) 
={1 \over {\pi \theta^2}}
\left[1-\left(1+{{\theta^2}\over {\theta_G^2}}\right)
\exp\left(-{{\theta^2}\over {\theta_G^2}}\right)\right],
\end{equation}
for $\theta < \theta_o$ and $Q=0$ elsewhere.
A nice property of this filter that should be noticed is that
it has less power on the inner region ($\theta <\theta_G$, see Fig.~1 of
\citet{2012MNRAS.425.2287H}) thus a ${\cal K }$-peak resulting from a cluster of
galaxies is less affected by contamination from cluster member galaxies.

In an actual computation, ${\cal K }$ is evaluated on regular
grid points with a grid spacing of 0.15~arcmin using equation
(\ref{eq:shear2kap}), but the integral in that equation is replaced
with the summation over galaxies taking into account the lensing weight.
The root-mean-square (RMS) noise coming from intrinsic ellipticity of
galaxies (which we call the galaxy shape noise and denote by
$\sigma_{\rm  shape}$) is evaluated on each grid point in the following
way: 
A ``noise ${\cal K }$ map'' is evaluated in the same way as the mass
map but the orientation of galaxies is randomized. 1000 realizations of
``noise ${\cal K}$ maps'' are produced, and the RMS of 1000 ${\cal K }$
values is evaluated on each grid, which we take as the local estimate of 
$\sigma_{\rm  shape}$. 
The signal-to-noise ratio of weak lensing mass map is defined by 
$S/N={\cal K }/\sigma_{\rm shape}$.
The filter parameters should be chosen so that the $S/N$ is maximized for
expected target clusters (i.e. $M>10^{14}h^{-1}M_\odot$ at $0.1<z<0.5$,
see \citet{2004MNRAS.350..893H}). 
We take $\theta_G=1.5$ arcmin and $\theta_o=15$ arcmin. 
We show in Fig.~\ref{fig:snmaps} the ${\cal K }$-$S/N$ maps for four
fields. 
Note that a ${\cal K }$-$S/N$ map is unaffected by the shear
multiplicative bias ($m$) because we used the same galaxy ellipticity
data for evaluating both the signal and noise, and thus the effect of
the multiplicative bias is canceled out.
Since in the following analysis we use only $S/N$ values, we did not
correct for the shear multiplicative bias.

On and around regions where no source galaxy is available due to imaging
data being affected by bright stars or large nearby galaxies, ${\cal K
}$ may not be accurately evaluated. 
We define ``data-region'', ``masked-region'' and ``edge-region'' by
using the distribution of source galaxies as follows:
First, for each grid, we check if there is a galaxy within 0.6
arcmin from the grid center. If there is no galaxy, then the grid is
flagged as ``no-galaxy''. After performing the procedure for all the
grids, all the ``no-galaxy'' grids plus all the grids within 0.6
arcmin from all the ``no-galaxy'' grids are defined as the ``masked-region''.
All the grids located within 1.5 arcmin (we take this value by setting
it equal to $\theta_G$) from
any of ``masked-region'' grids are defined as the ``edge-region''.
All the rest of grids are defined as the ``data-region''.
Areas of four fields are summarized in Table \ref{table:fields}.
The total survey area (``data-region''+ ``edge-region'') is 11.4
degree$^2$ with the ``edge-region'' accounting for 21 percent.

%
%
\begin{figure}
\begin{center}
\includegraphics[width=78mm]{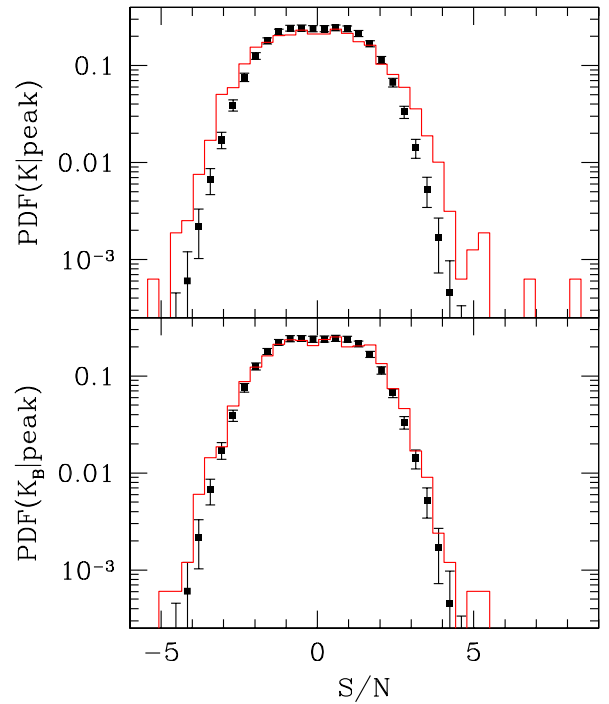}
\end{center}
\caption{The probability distribution function of positive- and
  negative-curvature peaks (peaks and troughs) is shown.
{\it Top-panel} is for $\cal K$ map, whereas {\it bottom-panel} is for
the B-mode  $\cal K$ map.
The histogram shows the results from observed $\cal K$ and ${\cal K}_B$
maps.
The points with error bars show the mean and RMS of PDF measured from
1000 random realizations of ``noise ${\cal K}$ maps'', which represents the PDF expected
for pure shape noise data.
\label{fig:fig5.eps}}
\end{figure}

Before looking into very high peaks, we examine statistical properties
of peaks paying a special attention to an impact of residual systematics
associated with data reduction and/or lensing shear measurement.
We computed the probability distribution function of positive- and
negative-curvature peaks (peaks and troughs) from the $\cal K$ maps, and
plotted in Fig.~\ref{fig:fig5.eps}.
We also computed the PDFs expected for pure shape noise data using 1000
random realizations of ``noise ${\cal K}$ maps'', and the mean and RMS
of 1000 PDFs are also plotted in the figure.
The upper panel of Fig.~\ref{fig:fig5.eps} shows the peak PDF
measured from $\cal K$ maps, where clear deviations from pure shape noise
PDF are seen.
The deviations originate mainly from three sources: (1) massive clusters
of galaxies that account for very high peaks (say $S/N>4.5$),
\citep{2002ApJ...580L..97M,2004MNRAS.350..893H} which we shall go
into details later,  
(2) cosmic large-scale structures which contribute to peaks at
$S/N \lesssim 4.5$ \citep{2010A&A...519A..23M,2010ApJ...709..286M}, and (3)
residual systematics.
In order to estimate the impact of the last one, we use the B-mode $\cal
K$ map (${\cal K}_B$), which is the same smoothed convergence map as
$\cal K$ but obtained from 45-degree rotated shear.
In the standard gravity theory, only a small amount of B-mode lensing
signal is produced \citep{2002A&A...389..729S,2009A&A...499...31H} which
should be negligible for the current analysis.
Since in the absence of lensing-induced B-mode signal, the ${\cal K}_B$-peak
PDF should follow the pure shape noise PDF, it can be used for a diagnosis of
residual systematics.
It is found from the bottom panel of Fig.~\ref{fig:fig5.eps} that the
${\cal K}_B$-peak PDF deviates from the pure shape noise PDF, indicating
existence of some residual systematics.
Although a cause of the excess ${\cal K}_B$-peaks is
unknown, it should be safer to consider that a similar amount of
non-lensing-originated peaks may exist in the ${\cal  K}$ maps. 
It can be also said from the ${\cal K}_B$ peak PDF that the probability
of having a non-lensing-induced peak with $S/N>5$ is very rare.
In addition to this fact, with a limited number of peaks with $S/N>5$, it is
possible to inspect all the individual high peaks.
Therefore we decided to use only the peaks with $S/N>5$ for our
cosmological study in the next section.
Understanding the origin of the excess ${\cal K}_B$ peak PDF must be
a very important subject for future studies of weak lensing cluster
counts using data from large surveys, however it is beyond the scope of
this paper and we leave it for a future work.

\subsection{High $S/N$ peaks}
\label{sec:peaks}

%
%
\begin{table*}
\caption{Summary of high peaks with $S/N>4.5$.}
\label{table:peaks}
\begin{tabular}{lccccllll}
\hline
Field - ID & R.A. & Dec. & Peak & flag$^{a}$ & 
\multicolumn{4}{c}{Matched known clusters ($\theta_{\rm sep}<2$ arcmin)} \\ 
{} & [deg] & [deg] & $S/N$ & {} & 
$\theta_{\rm sep}$$^{b}$ & $z_p$$^{c}$ & $z_{s,BCG}$$^{d}$ & Name and reference \\ 
\hline
XMM-LSS 1     &  36.6087 & $-5.0671$  & 4.7 & edge &
{} & {} & {} & {} \\
COSMOS 1      & 150.3017 &  1.5838  & 4.8 & - &  
$1.2'$ & 0.36 & {} & 
\#59 of $^{(1)}$\\
{} & {} & {}  & {} & {} & 
$1.5'$ & 0.242 & {} & 
NSC~J100113$+$013335$^{(2)}$\\
{} & {} & {}  & {} & {} & 
$1.5'$ & 0.37 & {} &
LSS~27 of $^{(3)}$\\
{} & {} & {}  & {} & {} & 
$1.9'$ & 0.3610 & 0.3638 & 
COSMOS~CL~J100112.4$+$013401$^{(4)}$ \\
COSMOS 2      & 149.4210 &  2.5585  & 5.1 & edge &
$0.5'$ & 0.3787 & {} &
WHL~J095742.9$+$023332$^{(5)}$\\
{} & {} & {}  & {} & {} & 
$1.0'$ & 0.17 & {} &
LSS 36 of $^{(3)}$\\
{} & {} & {} & {} & {} &  
$1.4'$ & 0.362 & 0.3733 & 
GMBCG~J149.40425$+$02.57381$^{(6)}$\\
{} & {} & {}  & {} & {} & 
$1.4'$ & 0.39 & {} &
LSS 26 of $^{(3)}$\\
Lockman-hole 1 & 161.6291 &  59.6266  & 7.6 & edge &
$0.7'$ & 0.2317 & 0.2282 &
WHL~J104625.5$+$593736$^{(5)}$ \\
{} & {} & {} & {} & {} &
$1.0'$ & 0.5672 & {} & 
SWIRE~CL~J104635.5$+$593732$^{(4)}$ \\
ELAIS-N1 1    & 244.2695 & 53.6936  & 4.6 & - &
$1.8'$ & 0.9061 & {} &
SWIRE~CL~J161710.9$+$534217$^{(4)}$\\
ELAIS-N1 2    & 242.8739 & 54.2766  & 5.2 & - &
$0.5'$ & 0.3794 & {} &
NSC~J161132$+$541649$^{(2)}$\\
{} & {} & {} & {} & {} &
$0.8'$ & {} & {} &
SWIRE~CL~J161130.1$+$541704$^{(4)}$\\
{} & {} & {} & {} & {} &
$0.9'$ & 0.3307 & {} &
WHL~J161135.9$+$541634$^{(5)}$ \\
ELAIS-N1 3    & 242.8169 & 55.5667  & 6.7 & - & 
$0.8'$ & 0.2246 & {} &
SWIRE~CL~J161118.8$+$553431$^{(4)}$\\
{} & {} & {}  & {} & {} & 
$1.1'$ & 0.2156 & {} &
WHL~J161121.1$+$553308$^{(5)}$\\
{} & {} & {}  & {} & {} & 
$1.8'$ & {} & {} &
ZwCl~1610.0$+$5543$^{(7)}$ \\
ELAIS-N1 4    & 242.6503 & 54.0947  & 5.1 & - & 
$1.0'$ & 0.3786 & 0.3370 & 
SWIRE~CL~J161045.2$+$540635$^{(4)}$\\
{} & {} & {}  & {} & {} &
$1.2'$ & 0.3618 & {} &
2XMM~J161040.5$+$540638$^{(8)}$\\ 
%
ELAIS-N1 5    & 241.6525 & 53.8947  & 5.3 & - & 
$0.2'$ & 0.3432 & 0.3505 & 
SWIRE~CL~J161045.2$+$540635$^{(4)}$ \\
{} & {} & {} & {} & {} & 
$0.7'$ &  0.348 & {} &
GMBCG~J241.67338$+$53.89618$^{(6)}$\\
ELAIS-N1 6    & 241.2632 & 55.6340  & 5.3 & - & 
$0.1'$ & 0.25385 & {} & 
MaxBCG~J241.26104$+$55.63254$^{(9)}$\\
{} & {} & {}  & {} & {} & 
$1.0'$ & 0.275 & {} &
WHL~J160503.4$+$553703$^{(5)}$\\
{} & {} & {}  & {} & {} & 
$1.2'$ & 0.2299 & {} &
NSC~J160459$+$553700$^{(2)}$\\
ELAIS-N1 7    & 241.1661 & 54.5305  & 8.2 & - & 
$0.2'$ & 0.2483 & {} &
WHL~J160439.4$+$543136$^{(5)}$ \\
\hline
\end{tabular}
$^{a}$ Flag for edge regions (see \S \ref{sec:kapmap} for the definition
of the edge region).\\
$^{b}$ Separation between the $\cal K$ peak and the cluster position
given in the references.\\
$^{c}$ Photometric redshift of clusters given in the references.\\
$^{d}$ Spectroscopic redshift of BCG cluster given in the references.\\
Reference list: $^{(1)}$\citet{2007ApJS..172..182F}, 
$^{(2)}$\citet{2003AJ....125.2064G}, 
$^{(3)}$\citet{2007ApJS..172..150S},
$^{(4)}$\citet{2011ApJ...734...68W},
$^{(5)}$\citet{2009ApJS..183..197W},
$^{(6)}$\citet{2010ApJS..191..254H},
$^{(7)}$\citet{1961cgcg.book.....Z},
$^{(8)}${\citet{2011A&A...534A.120T}},
$^{(9)}$\citet{2007ApJ...660..239K}.
\end{table*}

$\cal K$-peaks with $S/N>4.5$ are marked with open circles in
Fig.~\ref{fig:snmaps}, and are summarized in Table \ref{table:peaks}.
We searched known cluster database taken from a compilation by NASA/IPAC
Extragalactic Database (NED)\footnote{\tt http://ned.ipac.caltech.edu/}
for possible luminous baryonic counterparts within 2 arcmin from peak
positions, and matched clusters of galaxies are also presented in Table
\ref{table:peaks}.
Most of the matched known clusters were identified by galaxy
concentration in optical/infrared multi-band data and photometric
redshifts of the clusters were evaluated. For some cases the
spectroscopic redshifts of the brightest cluster galaxy (BCG) were
measured. 
It turns out that all but XMM-LSS 1, which has the peak height of
$S/N=4.7$ and is located in ``edge-region'', have matched known clusters.   
Although for some peaks, matched clusters are located at slightly
discrepant positions ($\theta_{\rm sep} > 1$ arcmin), all the peaks with
$S/N>5$ located in the ``data-region'' have a relatively well matched
cluster of galaxies ($\theta_{\rm sep} \leq 1$ arcmin).
Photometric redshifts of those closely matched counterparts are within the 
high lensing efficiency range ($0.1<z<0.5$), supporting an actual
correspondence between galaxy over-density and high weak lensing signal.
We thus conclude that the sample of $S/N>5$ peaks located in
the ``data-region'' are not contaminated by a false signal, and we use the
number counts of $N_{\rm peak}(S/N>5)=6$ in an effective area (``data-region''
only) of 9.0 degree$^2$ for the cosmological study in the next section.

%
%
\begin{figure}
\begin{center}
\includegraphics[width=78mm]{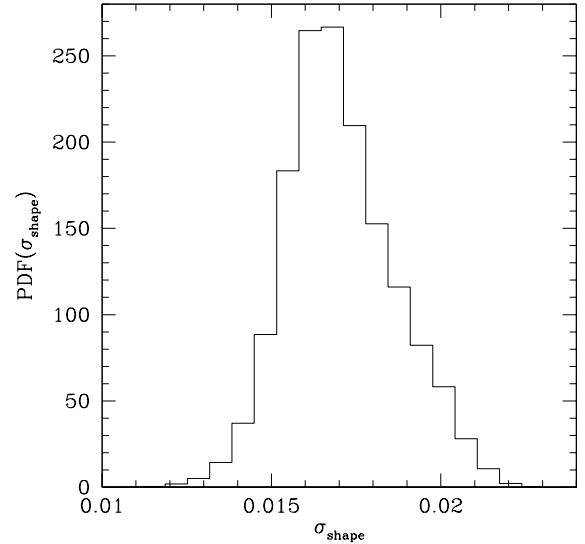}
\end{center}
\caption{The probability distribution of the shape noise $\sigma_{\rm
    shape}$ evaluated evaluated on each grid point using 1000 realizations of
``noise ${\cal K}$ maps'' as described in  \S \ref{sec:kapmap}.
The mean and RMS of this distribution are found to be $\langle \sigma_{\rm
    shape} \rangle =0.017$ and RMS$(\sigma_{\rm shape})=0.0016$, respectively.
\label{fig:fig6.eps}}
\end{figure}

\section{Constraints on the cosmological parameters and $M-c$ relation}
\label{sec:constraints}

Here we place constraints on the cosmological parameters ($\sigma_8$ and
$\Omega_m$) and the mass--concentration ($M-c$) relation using the
measured number counts of weak lensing clusters. 
To this end, we compare the measured counts with the theoretical model in
the standard $\chi^2$ method.
Below, we first describe the theoretical model of weak lensing cluster
counts, then next we evaluate the error in the measured number counts which
consist of the Poisson error and the cosmic variance, and
finally we present results.

\subsection{Theoretical model of weak lensing peak counts}
\label{sec:model}
We adopt the theoretical model of the weak lensing peak counts developed in
\citet{2012MNRAS.425.2287H}, which is based on halo models.
We refer the reader to the above reference for details of model and its
comparison with mock numerical simulations.
In short, the model is constructed under the assumption that high peaks
are dominated by lensing signals from single massive haloes, for which
 it is assumed that the density profile is described by the truncated NFW model
\citep{1997ApJ...490..493N,2009JCAP...01..015B} and mass function is given by the model of
\citet{2006ApJ...646..881W}.
Effects of intrinsic galaxy shape, the diversity of dark matter
distributions within haloes, and large-scale structures are included in
approximate approaches which were tested against the mock numerical
simulations.
Model parameters that should be specified for making the theoretical
prediction of the weak lensing cluster counts are the cosmological
parameters, $M-c$ relation of the dark matter haloes, the shape noise
$\sigma_{\rm shape}$, and the redshift distribution of source galaxies.
Below we will describe them in turn.

We treat $\Omega_m$ and $\sigma_8$ as free parameters, and fix the other
cosmological parameters as follows: Hubble parameter $h=0.7$,
the cosmological constant $\Omega_\Lambda = 1 -\Omega_m$, the baryon density
$\Omega_b = 0.051$, and the spectral index $n=0.96$, which are all
consistent with the recent CMB experiments
\citep{2013ApJS..208...19H,2014A&A...571A..16P}.

For the $M-c$ relation, we assume the following form,
\begin{equation}
\label{eq:M-c}
c_{\rm vir}(M_{\rm vir}, z) = c_0 
\left( {{M_{\rm vir}}\over 
{10^{12}h^{-1}M_{\odot}}}\right)^{-0.075}
(1+z)^{-0.7},
\end{equation}
which is motivated by the findings from dark matter N-body simulations
by
\citet{2008MNRAS.390L..64D,2008MNRAS.391.1940M,2011ApJ...740..102K,2012MNRAS.423.3018P},
but we treat the normalization as a free parameter.
\citet{2011ApJ...740..102K} found $c_0=9.6$ for the cosmological model
with $\Omega_m=0.27$ and $\sigma_8=0.82$.
However, it should be noted that the mass--concentration relationship
is known to be  dependent on the cosmological model
\citep{2008MNRAS.391.1940M,2009ApJ...707..354Z,2013ApJ...768..123K,2015ApJ...799..108D}.

The peak $\cal K$ value from a halo is computed by assuming the
tangential shear profile of a truncated NFW halo convolved with the
smoothing function, eq.~(\ref{eq:Q}).  Then it is converted to $S/N$ by 
specifying the shape noise, $\sigma_{\rm shape}$.
Since $\sigma_{\rm shape}$ depends on the number density of source
galaxies, it fluctuates depending on position.
As we have described in \S \ref{sec:kapmap}, we evaluated $\sigma_{\rm
  shape}$  on each grid of weak lensing mass maps. 
We computed the probability distribution of the
shape noise, which is presented in Fig.~\ref{fig:fig6.eps}.
The distribution is not very widely spread: its mean and RMS are found to be
$\langle \sigma_{\rm  shape} \rangle =0.017$ and RMS$(\sigma_{\rm
  shape})=0.0016$, respectively.
We take this effect into account in the following way,
\begin{equation}
\label{eq:Npeak_sigma_shape}
N_{\rm peak}(S/N) = \int d\sigma_{\rm shape}  P(\sigma_{\rm shape})
N_{\rm peak}(S/N; \sigma_{\rm shape}),
\end{equation}
where $N_{\rm peak}(S/N; \sigma_{\rm shape})$ is the theoretical
prediction of the weak lensing cluster number counts for a fixed
$\sigma_{\rm shape}$ and $P(\sigma_{\rm shape})$ is the probability
distribution function.
Note that $\sigma_{\rm shape}$ measured from the data is affected
  by the multiplicative bias in shear measurement (\S \ref{sec:shear}). 
In order to take it into account, we calibrate the shape noise RMS by  
$\tilde{\sigma}_{\rm shape} = \sigma_{\rm shape}/(1-\langle m \rangle)$,
and $P(\tilde{\sigma}_{\rm shape})$ with the measured distribution
function shown in  Fig.~\ref{fig:fig6.eps} is used to evaluate
eq.~{\ref{eq:Npeak_sigma_shape}}. 

%
%
\begin{figure}
\begin{center}
\includegraphics[width=78mm]{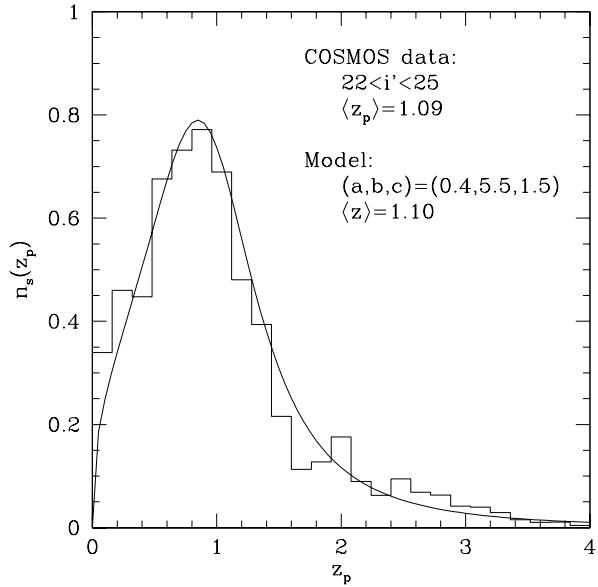}
\end{center}
\caption{Redshift distribution of galaxies used for the weak lensing
mass reconstruction derived adopting the COSMOS photometric redshift catalog
\citep{2009ApJ...690.1236I}. The solid curve shows the fitting model by
equation (8) with the parameters denoted in the panel. 
\label{fig:fig7.eps}}
\end{figure}

In order to infer the redshift distribution of source galaxies, we
utilized the COSMOS field data. 
We merged the source galaxy catalog used
for the weak lensing mass reconstruction with the public COSMOS
photometric redshift catalog \citep{2009ApJ...690.1236I},
and computed the redshift distribution by adopting the photometric
redshift.
The derived redshift distribution is  
presented in Fig.~\ref{fig:fig7.eps}. 
We fit the redshift distribution with the following function
\citep{2008A&A...479....9F}:
\begin{equation}
\label{eq:ns}
n_s(z) = A 
{{z^a + z^{ab}} \over {z^b + z^{c}}},
\end{equation}
where the normalization, $A$, was determined by imposing
$\int dz~n_s(z) = 1$ within the integration range $0 < z < 6$. 
We found that a parameter set $(a, b, c) = (0.4, 5.5, 1.5)$ gives a
reasonably good fit to the data, as shown in Fig.~\ref{fig:fig7.eps}.
The mean redshift of the model is 1.10 which is in very good agreement
with the value derived from the photometric redshift data (1.09).
In order to examine the variation in the redshift distribution due to
the difference in the exposure time, we divide the COSMOS field into two
subfields: one is the central region ($\sim 0.5$ degree$^2$) where the
exposure time is longer than 3,600 sec (with a median of 4,920 sec), and
the other is the outer region where the exposure time is shorter than
3,600 sec (but mostly 2,400 sec) \citep[see][for variation of the
  limiting magnitude]{2007ApJS..172....9T}.
We computed the redshift distributions of those two samples separately,
and found that the distributions are nearly indistinguishable, with mean
redshifts of 1.093 and 1.090 for the inner-deep and outer-shallow samples, respectively. 
This may be partly due to our selection criteria that $i'<25$ [AB mag],  
detection $S/N>8$ and FWHM larger than the stellar FWHM, which
may result in rejection of small and faint distant galaxies.
According to this finding and the fact that the redshift information for
the other three fields is not available, we do not consider a possible
field-to-field variation in the redshift distribution but use
eq.~(\ref{eq:ns}) for computation of the theoretical prediction.

%
%
\begin{figure}
\begin{center}
\includegraphics[width=78mm]{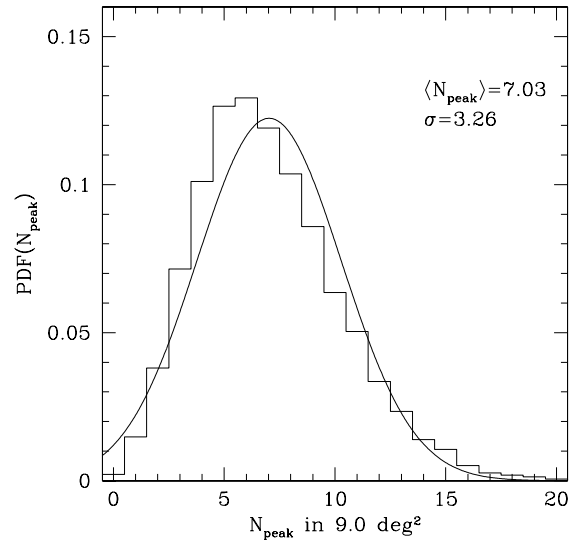}
\end{center}
\caption{The probability distribution function of the number counts of
  peaks with $S/N>5$ found in 9.0 degree$^2$ fields.
The distribution is derived from 10,000 realizations randomly taken from the 
full-sky mock weak lensing data.
The mean and RMS of the distribution is 7.03 and 3.26, respectively.
\label{fig:fig8.eps}}
\end{figure}

\subsection{Uncertainty in the number counts}
\label{sec:error}

In order to evaluate the uncertainty in the weak lensing cluster counts, 
we utilize mock weak lensing data generated from
full-sky gravitational lensing ray-tracing simulations.
Details of the simulation are given in Shirasaki, Hamana \& Yoshida
(2015 in prep.), here we describe only aspects that are directly relevant for
this work.
We adopted the cosmological model with $\Omega_m = 0.278$,
$\Omega_\Lambda = 0.722$, $\sigma_8 = 0.82$, $h = 0.70$ and $n =
0.97$. These parameters are consistent with the WMAP nine-year results 
\citep{2013ApJS..208...19H}.
Gravitational lensing ray-tracing was performed through the dark matter
distribution generated by the $N$-body simulations.
In order to cover full-sky up to $z>1$, we combined, in a nested structure,
8 simulation boxes of different sizes with box side length from
450$h^{-1}$Mpc to 3,600$h^{-1}$Mpc.
The $N$-body simulations were run with $2048^3$ dark matter particles
using the parallel Tree-Particle Mesh code {\it Gadget2}
\citep{2005MNRAS.364.1105S}. 
Light ray paths and the magnification matrix were evaluated by the
multiple-plane ray-tracing algorithm of \citet{2009A&A...497..335T} and
\citet{2013MNRAS.435..115B}.
We used HEALPix pixelization of a sphere \citep{2005ApJ...622..759G}. 
The angular resolution parameter $nside$ was set to 8192. 
The corresponding angular resolution is $0.43$ arcmin.
We adopted a fixed source redshift of $z_s=1.09$.
On each pixel center, we computed the lensing shear.
We added a random galaxy shape noise drawn from the truncated
two-dimensional Gaussian to the lensing shear data 
\citep{2012MNRAS.425.2287H}.
The $\sigma$ of the Gaussian function was determined so that the shape
noise in the resultant $\cal K$ map became equal
to the mean observed value (after calibration with the
  multiplicative bias), that is $\tilde{\sigma}_{\rm shape}=
\sigma_{\rm shape}/(1-\langle m \rangle)=0.018$.
We adopted the same smoothing function as the actual measurement,
eq.~(\ref{eq:Q}), and computed $\cal K$ from the noise-added shear data.
Since we used the dark matter distribution from cosmological $N$-body
simulations, the influence of large-scale structures on weak lensing
cluster counts should be properly included. 

Having generated the mock $\cal K$ data, we count the number of peaks
with the threshold of $S/N>5$ in a contiguous 9.0 degree$^2$ region,
which is randomly taken from the full-sky simulation data.
We repeat this counting 10,000 times, allowing overlap in regions between
different realizations.
The probability distribution function of the resulting number counts is
shown in Fig.~\ref{fig:fig8.eps}.
The mean and RMS of the distribution was found to be 7.03 and 3.26,
respectively. 
Note that the solid curve in Fig.~\ref{fig:fig8.eps}
represents the Gaussian distribution with the same mean and
RMS, demonstrating that the measured distribution can be reasonably
approximated by the Gaussian distribution.
We may safely assume that the RMS consists of Poisson error and cosmic
variance. Taking into account the difference between the observed number
counts of 6 and the mean number counts of 7.03, we adopt $\sigma_N =
3.1$ for the error in the observed weak lensing cluster counts, and thus
we finally have $N_{\rm peak} = 6 \pm 3.1$.

%
%
\begin{figure}
\begin{center}
\includegraphics[height=84mm,angle=-90]{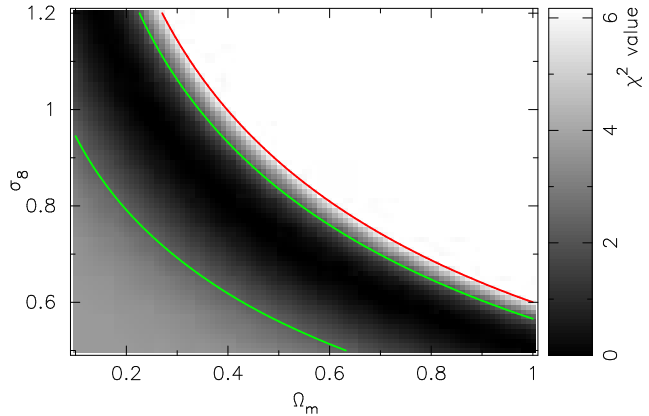}
\end{center}
\caption{Constraints on $\Omega_m$ and $\sigma_8$ derived from 
the weak lensing cluster counts. Here the normalization of the $M-c$
relation is assumed to be $c_0=9.6$ \citep{2011ApJ...740..102K}.
Gray scale shows the $\chi^2$ with the inner green and outer red (upper
limit only) contours indicating 68.3\% ($\chi^2=2.30$), 95.4\%
($\chi^2=6.17$) confidence levels, respectively.
\label{fig:fig9.eps}}
\end{figure}

%
%
\begin{figure}
\begin{center}
\includegraphics[height=84mm,angle=-90]{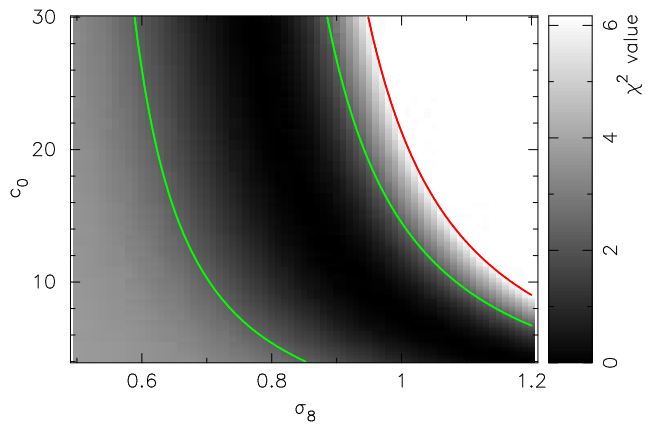}
\end{center}
\caption{Same as Fig.~\ref{fig:fig9.eps} but on the
  $\sigma_8$ and $c_0$. Here the density parameter was fixed with
  $\Omega_m =0.278$.
\label{fig:fig10.eps}}
\end{figure}

%
%
\begin{figure}
\begin{center}
\includegraphics[width=72mm]{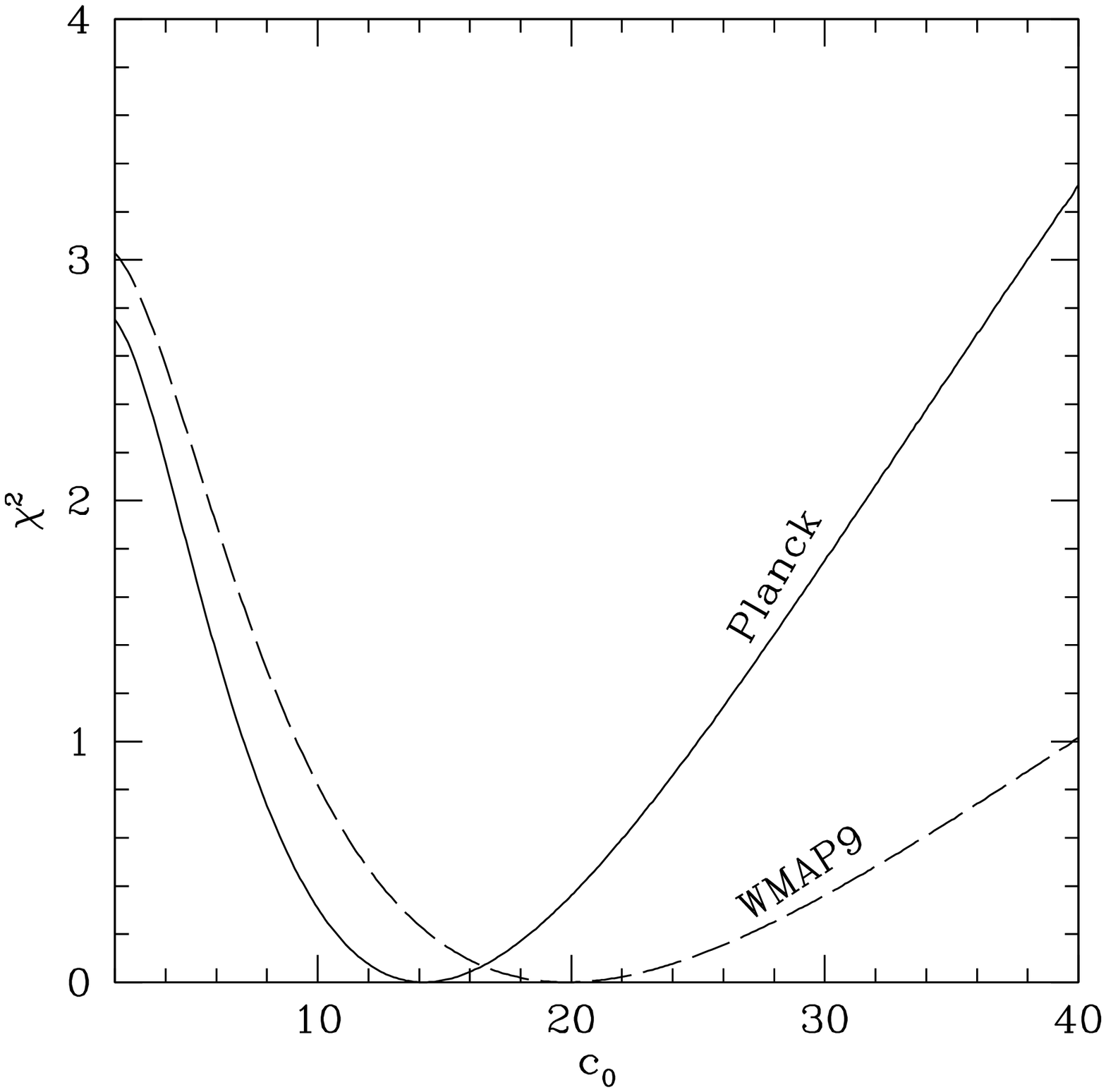}
\end{center}
\caption{Constraint on a single parameter $c_0$ derived from 
the weak lensing cluster counts. 
Here $\Omega_m$ and $\sigma_8$ are fixed to be the best fitting model of
either the Planck \citep[solid curve][]{2014A&A...571A..16P} or WMAP
9-year \citep[dashed curve;][]{2013ApJS..208...19H}.
\label{fig:fig11.eps}}
\end{figure}

\subsection{Results}
\label{sec:results}

In order to place constraints on cosmological
parameters  and the $M-c$ relation using weak lensing cluster counts, 
we compare the measured counts with the theoretical
prediction utilizing the standard $\chi^2$ method.
Note that the resulting $\chi^2$ value does not 
represent the goodness-of-fit, as we have only one observed value, 
$N_{\rm peak} = 6 \pm 3.1$.
Thus we use the $\chi^2$ value only for placing a confidence region on 
parameter spaces, adopting confidence levels assuming a normal
distribution. 
We treat $\Omega_m$, $\sigma_8$ and $c_0$ as free parameters.
Constraints obtained from the weak lensing cluster counts
have degeneracy between those parameters \citep{2014JCAP...08..063M,2014arXiv1409.5601C}.
However, since we have only one observed value with poor statistics, we
do not explore the three-parameter space, but we examine two-parameter space,
fixing one remaining parameter. 

Figure \ref{fig:fig9.eps} shows the derived
constraint on $\Omega_m-\sigma_8$ plane. 
Here we did not treat $c_0$
as a fitting parameter but adopted the value ($c_0=9.6$) found from dark
matter $N$-body simulations by \citet{2011ApJ...740..102K} in which the
$\Lambda$CDM model with $\Omega_m=0.27$ and $\sigma_8=0.82$ was
employed.
Therefore the constraint mainly comes from the cosmology dependence of
the dark matter halo mass function with a minor contribution from the
cosmology dependent lensing efficiency.
Accordingly, the constraint exhibits the $\Omega_m-\sigma_8$
degeneracy in a similar fashion to X-ray cluster counts
\citep{2001ApJ...561...13B,2004ApJ...609..603H,2009ApJ...692.1060V,2014A&A...570A..31B}.

Figure \ref{fig:fig10.eps} shows the derived
constraint on the $\sigma_8-c_0$ plane, where the density parameter was
fixed with $\Omega_m=0.278$. 
The degeneracy in the $\sigma_8-c_0$ space results from two
relationships: one is increasing dark matter halo number density for
larger $\sigma_8$,  and the other is higher $S/N$-peak for larger
$c_0$. Both of them result in larger weak lensing cluster counts.
The upper limit on $c_0$ is hardly placed, because of the degeneracy in
$\sigma_8-c_0$ and, for a given smoothing
kernel, an excessively concentrated halo compared with the smoothing scale
causes little variation in the peak height \citep{2004MNRAS.350..893H}. 
Although the current constraint does not even place a useful upper/lower
limit on those parameters due to the poor statistics, this demonstrates
that weak lensing cluster counts can be a useful probe of the
cosmic structure formation scenario.
The same argument, but based on a slightly different methodology, was
already made by \cite{2014arXiv1412.3683L}.

Finally, for a demonstrative purpose, we present in Figure
\ref{fig:fig11.eps}, the constraint on a single parameter $c_0$
for two representative cosmological models: the Planck
($\Omega_m=0.315$, $\sigma_8=0.83$) \citep{2014A&A...571A..16P} and
WMAP 9-year ($\Omega_m=0.278$, $\sigma_8=0.82$)
\citep{2013ApJS..208...19H}.
It is obvious that the measured weak lensing cluster counts are
consistent with the currently popular $\Lambda$CDM model.
Specifically, the result from $N$-body simulations by
\citet{2011ApJ...740..102K} ($c_0=9.6$) is well within the 68.3\%
confidence region of both the Planck and WMAP9 cosmologies.

\section{Summary and discussions}
\label{sec:summary}

We presented the results of weak lensing cluster counts obtained from 11
degree$^2$ Subaru/SuprimeCam data.
For the first time, we used the weak lensing cluster counts to place
cosmological parameters and the mass--concentration relationship.
In doing this, we explored, in an experimental manner, the capability
and current issues of weak lensing cluster counts for cosmological
studies. 

The data were selected under the conditions that the exposure time
longer than 1,800~sec along with the seeing condition of ${\rm FWHM} <
0.7$~arcsec. 
These selection criteria result in the effective number density of galaxies
usable for weak lensing shape measurement of $n_g \gtrsim 20$  arcmin$^{-2}$,
which is about twice as large as those of
previous works dealing with weak lensing peak statistics, namely, the
CFHTLenS by \citet{2014arXiv1412.0757L} and CFHT Stripe-82 Survey by
\citet{2014arXiv1412.3683L}. 
Since the shape noise in the $\cal K$ map scales as $\sigma_e \propto
n_g^{-1/2}$, higher galaxy density lifts up the peak height of a lensing
signal of massive clusters.
This allows us to adopt a high threshold $S/N$, in our case $S/N>5$.
With high $S/N$, the contamination rate due to false signals is
small, which enables us to construct a weak lensing-selected cluster
sample. 

We used {\it lensfit} for weak lensing shear measurement. 
We examined the additive bias in the measured galaxy ellipticities by
evaluating weighted mean ellipticities as a function of galaxy size
and magnitude.
We found non-zero additive bias for both $e_1$ and $e_2$, which were
found to be mainly caused by faint and small galaxies, as shown in
Fig.~\ref{fig:fig3}. 
This is in marked contrast to the result of CFHTLenS in which 
small and bright galaxies were found to be a cause of
additive bias in $e_2$ \citep{2012MNRAS.427..146H}. 
The cause of these non-zero biases is unclear, but those different trends
for different data sets may suggest that it may be caused by some
instrumental or data-dependent effect.
We evaluated the amplitude of the multiplicative bias, adopting the
empirical fitting function derived from CFHTLenS image simulations
\citep{2013MNRAS.429.2858M}, and found the weighted average value of
$\langle m \rangle= -0.059$.
This does not have influence on the measured peak-$S/N$, as both the signal and noise
RMS are affected in the same way, and thus those are canceled out.
However, this must be properly taken into account in making the
theoretical prediction.
Since in the current study we considered the statistical quantity
(peak counts), we adopted a statistical correction scheme that
calibrating the shape noise RMS with the average bias factor
$(1-\langle m \rangle)$.

We found 6 peaks with $S/N>5$.
For all the peaks, previously identified clusters of galaxies are
matched within a separation of $\theta_{\rm sep} < 1$ arcmin.
Most of the matched clusters were identified as a galaxy concentration
in the sky and in multi-band color space of optical/infrared data.
This provides us with a firm inspection of the purity of the weak lensing
cluster sample.
In fact, optical multi-band images are well suited to
cross-identification, as the sensitivity of weak lensing mass map to
cluster finding is most effective for clusters within the redshift range
$0.1 \lesssim z \lesssim 0.6$, where the red-sequence of cluster galaxies
is located in optical color space.
Thus, ongoing optical multi-band galaxy surveys with a single band
observation being optimized for weak lensing shape measurement, namely
the Kilo-Degree Survey (KiDS)\footnote{\tt http://kids.strw.leidenuniv.nl},  
Dark Energy Survey (DES)\footnote{\tt http://www.darkenergysurvey.org}, 
and Hyper-SuprimeCam (HSC) Survey \citep{2012SPIE.8446E..0ZM},
may provide self-contained datasets for weak lensing cluster finding.
Also, photometric redshifts will be estimated from multi-band data
for each galaxy, which enable the source galaxy
redshift distributions to be accurately determined.

We evaluated the statistical error in the weak lensing cluster counts
using mock weak lensing data generated from full-sky ray-tracing simulations, 
and found $N_{\rm peak}= 6\pm 3.1$.
This error consists of the Poisson noise and the cosmic variance, with
a larger contribution from the former.

We compared the measured weak lensing cluster counts with the
theoretical model developed by
\citet{2004MNRAS.350..893H,2012MNRAS.425.2287H}, in which 
the effects of intrinsic galaxy shape, the diversity of dark matter
distributions within haloes, and large-scale structures are included.
Utilizing the standard $\chi^2$ method, we placed constraints on the
$\Omega_m-\sigma_8$ plane, which was found to be consistent
with currently standard $\Lambda$CDM models such as Planck model
\citep{2014A&A...571A..16P} and WMAP 9-year model \citep{2013ApJS..208...19H}, 
though the constraints are much broader than those CMB experiments due
to the poor statistics ($N_{\rm  peak}= 6\pm 3.1$).
It was demonstrated that the weak lensing cluster counts may place a
unique constraint on $\sigma_8-c_0$ plane.

Finally we discuss prospects for ongoing/future surveys, taking the DES
and HSC surveys as examples.
Even taking into account a lower usable galaxy number
density for those surveys, due to the shallower survey depth than this study,
it may be feasible to obtain $N_{\rm peak} \sim 400$ clusters with $S/N
\gtrsim 5$  
(which can be achieved with $\sim 0.3$ peak/degree$^2$ for HSC survey of
1,400 degree$^2$, and $\sim 0.08$ peak/degree$^2$ for DES of 5,000
degree$^2$). In that case, the error is dominated by the Poisson error
and the expected fractional error will be $N_{\rm peak}^{-1/2} \sim 0.05$.
This is 10 times better than the current study, $3.1/6 \sim 0.5$.
Therefore confidence regions become smaller and thus tighter
constraints will be obtained; taking Figure
\ref{fig:fig11.eps} as an illustrative example, the
corresponding 68.3\% confidence level becomes $\chi^2=0.1$. 
This demonstrates that this methodology will be a useful
probe of the cosmology and $M-c$ relation, though, in reality,
marginalization over other parameters and degeneracy among
parameters weaken constraints. 
More quantitative investigations of the capability of this method based
on a Fisher analysis were made by
\citet{2014arXiv1409.5601C,2014JCAP...08..063M}. 
Having found that tight cosmological constraints can be achieved using
the significantly improved statistics expected for ongoing/future surveys, 
it is necessary to check whether the theoretical model is
correspondingly accurate, which is beyond the scope of
this paper and is left for a future work. 

\bigskip

We thank Y. Okura, M. Takada, M. Oguri and S. Miyazaki for useful
discussions/comments. 
We are grateful to R. Takahashi and M. Shirasaki for assistance with
running the $N$-body simulations for the full-sky gravitational lensing
simulation.
We would like to thank Matthew R. Becker for making the source codes of
CALCLENS publicly available, and HEALPix team for HEALPix software
publicity available.
This paper makes use of software developed for the Large Synoptic Survey
Telescope. We thank the LSST Project for making their code available as
free software at {\tt http://dm.lsstcorp.org}. 
We would like to thank HSC Data Analysis Software Team for their effort
to develop {\it hscPipe} software suite.
TH and LM thank the Aspen Center for Physics for their warm hospitality,
where this work was partly done.
Numerical computations in this paper were in part carried out Cray XC30
at Center for Computational Astrophysics, National Astronomical
Observatory of Japan. 
This work is based in part on data collected at Subaru Telescope and
obtained from the SMOKA, which is operated by the Astronomy Data Center,
National Astronomical Observatory of Japan.
This work is supported in part by Grant-in-Aid for
Scientific Research from the JSPS Promotion of Science
(23540324).


\appendix

\section{Method of conversion from SIP to TPV}
\label{appendix:sip2tpv}

As described in \S \ref{sec:dataanalysis}, the instrumental distortion
of images was corrected in the process of mosaic stacking along with
astrometric calibration of CCD data using {\it hscPipe} in
which the relationship between pixel coordinates and world coordinates
systems (WCSs) is represented by SIP convention \citep{2005ASPC..347..491S}. 
In {\it lensfit}, on the other hand, TPV
convention\footnote{http://fits.gsfc.nasa.gov/registry/tpvwcs/tpv.html}
is implemented. Here we describe the method for conversion of WCSs from
SIP to TPV. 

\subsection{SIP convention}
\label{subsec:sip}

First, we summarize the SIP convention following
\citet{2005ASPC..347..491S}. 
Let $u,v$ be relative pixel coordinates on a detector device with origin
at {\tt CRPIX1}, {\tt CRPIX2}, and $x',y'$ be distortion corrected
``intermediate world coordinates'' in degree with origin at {\tt
  CRVAL1}, {\tt CRVAL2}.
Then
\begin{equation}
\label{eq:sip1}
\left(
\begin{array}{c}
x'\\
y'
\end{array}
\right)
=
\left(
\begin{array}{cc}
\cd{1}{1} &  \cd{1}{2} \\
\cd{2}{1} &  \cd{2}{2}
\end{array}
\right)
\left(
\begin{array}{c}
u + f(u,v)\\
v + g(u,v) 
\end{array}
\right),
\end{equation}
where $\cd{i}{j}$ matrix encodes scaling, rotation and skew, and
$f(u,v)$ and $g(u,v)$ are polynomial functions that represent 
distortion, given by 
\begin{eqnarray}
\label{eq:fg-AB}
f(u,v) &=& \sum_{i,j} \A{i}{j} u^i v^j, \quad i+j \leq {\rm \tt A\_ORDER},\\
g(u,v) &=& \sum_{i,j} \B{i}{j} u^i v^j , \quad i+j \leq {\rm \tt B\_ORDER}.
\end{eqnarray}
In the standard SIP convention, {\tt A\_ORDER} and {\tt B\_ORDER} are
allowed to range from 2 to 9.

\subsection{TPV convention}
\label{subsec:tpv}

Next, we summarize the TPV convention, in which relative pixel
coordinates ($u,v$) are first converted to ``distorted world
coordinates'' $(x,y)$ by
\begin{equation}
\label{eq:tpv1}
\left(
\begin{array}{c}
x\\
y
\end{array}
\right)
=
\left(
\begin{array}{cc}
\cd{1}{1} &  \cd{1}{2} \\
\cd{2}{1} &  \cd{2}{2}
\end{array}
\right)
\left(
\begin{array}{c}
u \\
v  
\end{array}
\right).
\end{equation}
Then ``intermediate world coordinates'' $(x',y')$ can be given by
polynomial functions of $x$ and $y$ as,
\begin{eqnarray}
\label{eq:x'y'-xy}
x' &=& \pv{1}{0} 
+ \pv{1}{1}x + \pv{1}{2}y + \pv{1}{3}r  \nonumber\\
&& + \pv{1}{4}x^2 + \pv{1}{5}xy + \pv{1}{6}y^2 \nonumber\\
&& + \pv{1}{7}x^3 + \pv{1}{8}x^2y + \pv{1}{9}xy^2 \nonumber\\
&& + \pv{1}{10}y^3 + \pv{1}{11}r^3+\cdots, \\
y' &=& \pv{2}{0} 
+ \pv{2}{1}y + \pv{2}{2}x + \pv{2}{3}r  \nonumber\\
&& + \pv{2}{4}y^2 + \pv{2}{5}xy + \pv{2}{6}x^2 \nonumber\\
&& + \pv{2}{7}y^3 + \pv{2}{8}xy^2 + \pv{2}{9}x^2y \nonumber\\
&& + \pv{2}{10}x^3 + \pv{2}{11}r^3+\cdots,
\end{eqnarray}
where $r=\sqrt{x^2+y^2}$, and the TPV polynomial functions have
odd-power terms of $r$. 
The polynomial order of TPV is defined up to 7th, thus there are $2\times
40$ PV$i$\_$j$ coefficients including $2\times 4$ odd $r^n$-terms.

\subsection{Conversion from SIP to TPV}
\label{subsec:sip2tpv}

We derive conversion equations from SIP coefficients to TPV
coefficients.
To do so, we limit the polynomial order of SIP ({\tt A\_ORDER} and {\tt
  B\_ORDER}) being equal to or less than 7th, because that of TPV is
defined up to 7th.

We can set PV$i$\_$j$ coefficients for
$r^n$-terms\footnote{Those are $\pv{1}{3}$, $\pv{1}{8}$,  $\pv{1}{23}$,
  $\pv{1}{39}$, $\pv{2}{3}$, $\pv{2}{8}$,  $\pv{2}{23}$ and $\pv{2}{39}$.} zero,
because there are no corresponding terms in SIP polynomials.
Also, we can safely use the same origins of $(u,v)$ and $(x',y')$
coordinates for TPV as those for SIP, because we derive the same
distortion polynomial models as those of SIP but in the different (TPV) 
convention (thus the same {\tt CRPIX1}, {\tt CRPIX2}, {\tt CRVAL1} and
{\tt CRVAL2} can be used). 
Then, we can set $\pv{1}{0} = \pv{2}{0} =0$ as there are no
corresponding terms in SIP.
In addition, utilizing redundant degree of freedom in TPV convention, we
can safely take the same $\cd{i}{j}$ matrix as one of SIP.

Equating the SIP and TPV coefficients of each $u^i v^j$ terms by
order-by-order, we have the following simultaneous equations:
For the 1st order,
\begin{eqnarray}
\label{eq:sip2tpv1st}
\left(
\begin{array}{c}
\cd{1}{1}\\
\cd{1}{2}
\end{array}
\right)
&=&
\left(
\begin{array}{cc}
\cd{1}{1} &  \cd{2}{1} \\
\cd{1}{2} &  \cd{2}{2}
\end{array}
\right)
\left(
\begin{array}{c}
\pv{1}{1} \\
\pv{1}{2}
\end{array}
\right),\\
\left(
\begin{array}{c}
\cd{2}{1}\\
\cd{2}{2}
\end{array}
\right)
&=&
\left(
\begin{array}{cc}
\cd{2}{1} &  \cd{1}{1} \\
\cd{2}{2} &  \cd{1}{2}
\end{array}
\right)
\left(
\begin{array}{c}
\pv{2}{1} \\
\pv{2}{2}
\end{array}
\right),
\end{eqnarray}
which can be inverted and can be solved as,
\begin{equation}
\label{eq:sip2tpv1st-sol}
\left(
\begin{array}{c}
\pv{1}{1} \\
\pv{1}{2}
\end{array}
\right)
=
\left(
\begin{array}{c}
\pv{2}{1} \\
\pv{2}{2}
\end{array}
\right)
=
\left(
\begin{array}{c}
1\\
0
\end{array}
\right).
\end{equation}
These are a natural consequence of distortion polynomial functions of
SIP convention being defined only for 2nd and higher order terms (see
eqs.~(\ref{eq:sip1}) and (\ref{eq:fg-AB})).
For the 2nd order, we have
\begin{eqnarray}
\label{eq:sip2tpv2ndA}
&&\left(
\begin{array}{c}
\A{2}{0} \cd{1}{1} + \B{2}{0} \cd{1}{2}\\
\A{1}{1} \cd{1}{1} + \B{1}{1} \cd{1}{2}\\
\A{0}{2} \cd{1}{1} + \B{0}{2} \cd{1}{2}
\end{array}
\right)=\nonumber\\
&&
{\footnotesize
\left(
\begin{array}{ccc}
\cd{1}{1}^2 &  \cd{1}{1} \cd{2}{1} & \cd{2}{1}^2 \\
2\cd{1}{1}\cd{1}{2} &  \cd{1}{1} \cd{2}{2} + \cd{1}{2} \cd{2}{1} &
2\cd{2}{1}\cd{2}{2}\\
\cd{1}{2}^2 &  \cd{1}{2} \cd{2}{2} & \cd{2}{2}^2 \\
\end{array}
\right) 
}\nonumber \\
&&\times
\left(
\begin{array}{c}
\pv{1}{4} \\
\pv{1}{5} \\
\pv{1}{6}
\end{array}
\right),
\end{eqnarray}
and
\begin{eqnarray}
\label{eq:sip2tpv2ndB}
&&\left(
\begin{array}{c}
\A{2}{0} \cd{2}{1} + \B{2}{0} \cd{2}{2}\\
\A{1}{1} \cd{2}{1} + \B{1}{1} \cd{2}{2}\\
\A{0}{2} \cd{2}{1} + \B{0}{2} \cd{2}{2}
\end{array}
\right)=\nonumber\\
&&
{\footnotesize
\left(
\begin{array}{ccc}
\cd{2}{1}^2 &  \cd{2}{1} \cd{1}{1} & \cd{1}{1}^2 \\
2\cd{2}{1}\cd{2}{2} &  \cd{2}{1} \cd{1}{2} + \cd{2}{2} \cd{1}{1} &
2\cd{1}{1}\cd{1}{2}\\
\cd{2}{2}^2 &  \cd{2}{2} \cd{1}{2} & \cd{1}{2}^2 \\
\end{array}
\right) 
}\nonumber \\
&&\times
\left(
\begin{array}{c}
\pv{2}{4} \\
\pv{2}{5} \\
\pv{2}{6}
\end{array}
\right).
\end{eqnarray}
In the same manner, one can derive simultaneous equations in matrix
form for higher order terms.
Therefore, TPV coefficients, $\pv{i}{j}$, can be obtained by solving
those simultaneous equations using the standard matrix inversion method.
In an actual computation, we performed matrix inversion and
multiplication numerically.

\end{document}